\documentclass[twoside,12pt]{article}
\usepackage{epsfig}
\usepackage{color}

\def\Journal#1#2#3#4{{#1} {#2} (#4) #3 }
\def\NCA{{\em Nuovo Cimento} A}

\def\NPA{{\em Nucl. Phys.} A}

\def\PRO{{\em Prog. Theor. Phys.}}
\def\NPB{{\em Nucl. Phys.} B}

\def\PLB{{\em Phys. Lett.} B}

\def\PRL{\em Phys. Rev. Lett.}

\def\PREP{\em Phys. Rep.}

\def\PRD{{\em Phys. Rev.} D}
\def\PRC{{\em Phys. Rev.} C}

\def\ZPC{{\em Z. Phys.} C}

\def\ANNP{\em Ann. Phys. (N.Y.)}
\def\RMP{{\em Rev. Mod. Phys.}}

\def\JPG{\em J. Phys. G: Nucl. Part. Phys.} 

\newcommand{\be}{\begin{equation}}
\newcommand{\ee}{\end{equation}}
\newcommand{\bea}{\begin{eqnarray}}
\newcommand{\eea}{\end{eqnarray}}

\begin{document}

\title{\vspace{1cm} Shannon Information Entropy in Heavy-ion Collisions}
\author{Chun-Wang\ Ma
\thanks{Email: machunwang@126.com}
\\
Institute of Particle and Nuclear Physics, Henan Normal University, China\\
Yu-Gang\ Ma
\thanks{Email: ygma@sinap.ac.cn}
\\
Shanghai Institute of Applied Physics, Chinese Academy of Sciences, China\\
}
\maketitle
\begin{abstract}
The general idea of information entropy provided by C.E. Shannon ``hangs over everything we do'' and can be applied to a great variety of problems once the connection between a distribution and the quantities of interest is found. The Shannon information entropy essentially quantify the information of a quantity with its specific distribution, for which the information entropy based methods have been deeply developed in many scientific areas including physics. The dynamical properties of heavy-ion collisions (HICs) process make it difficult and complex to study the nuclear matter and its evolution, for which Shannon information entropy theory can provide new methods and observables to understand the physical phenomena both theoretically and experimentally. To better understand the processes of HICs, the main characteristics of typical models, including the quantum molecular dynamics models, thermodynamics models, and statistical models, etc, are briefly introduced. The typical applications of Shannon information theory in HICs are collected, which cover the chaotic behavior in branching process of hadron collisions, the liquid-gas phase transition in HICs, and the isobaric difference scaling phenomenon for intermediate mass fragments produced in HICs of neutron-rich systems. Even though the present applications in heavy-ion collision physics are still relatively simple, it would shed light on key questions we are seeking for. It is suggested to further develop the information entropy methods in nuclear reactions models, as well as to develop new analysis methods to study the properties of nuclear matters in HICs, especially the evolution of dynamics system.
\end{abstract}

\tableofcontents
\section{Introduction\label{sec:introd}}

In his fundamental articles constructing the information entropy theory in 1948 \cite{CESh1}, the father of the information age, C.E. Shannon ``transformed information from a vague idea into a precise concept that underlies the digital revolution'' \cite{ComShnn}. Based on the initiative concept of information entropy, the mathematical descriptions for information transfer in communication unify a diverse bunch of technologies. Shannon's fundamental theorem, as commented by R. Lucky \cite{ComShnn}, has served as an ideal and a challenge for succeeding generations, which also ``hangs over everything we do'' and inspired the development of ``all our modern error-correcting codes and data-compression algorithms''. The dimension of information, which underlies any distribution of data, yields an upper bound on the compression rate of any variable in a given distribution \cite{ComShnn1}. The information entropy, which is defined as the measurement of the degree of chaos, denotes the chaoticity of a system. These ideas of information entropy are very general, and can be applied to a great variety of problems if one finds some distributions of the system connected to quantities of interest \cite{Soli01}. The theorems proved that the Shannon information entropy is equivalent to information chaoticity carried by message, which means that the larger information entropy of a system, the larger chaoticity it contains.

After Shannon opened the era of information, information entropy has found important applications in the fields of cybernetics, probability theory, number theory, astrophysics, life science, social system, stocks analysis, and so on. The information entropy theory is applied in characterizing dynamical systems between regular, chaotic and purely random evolution. Among these problems, three categories of communication systems can be dealt with: discrete, continuous, and discrete/continous mixed \cite{CESh1}. Though the definition of Shannon information entropy shares the form of entropy defined in thermodynamics, it avoids the limitation required for a physical system by thermodynamics, and measures the chaotic evolution and loss of information in dynamical evolution of system. This advantage makes it a good tool to study the evolution of dynamical processes of physical systems.

\begin{figure}[htbp]
\begin{center}
\begin{minipage}[t]{16 cm}
\epsfig{file=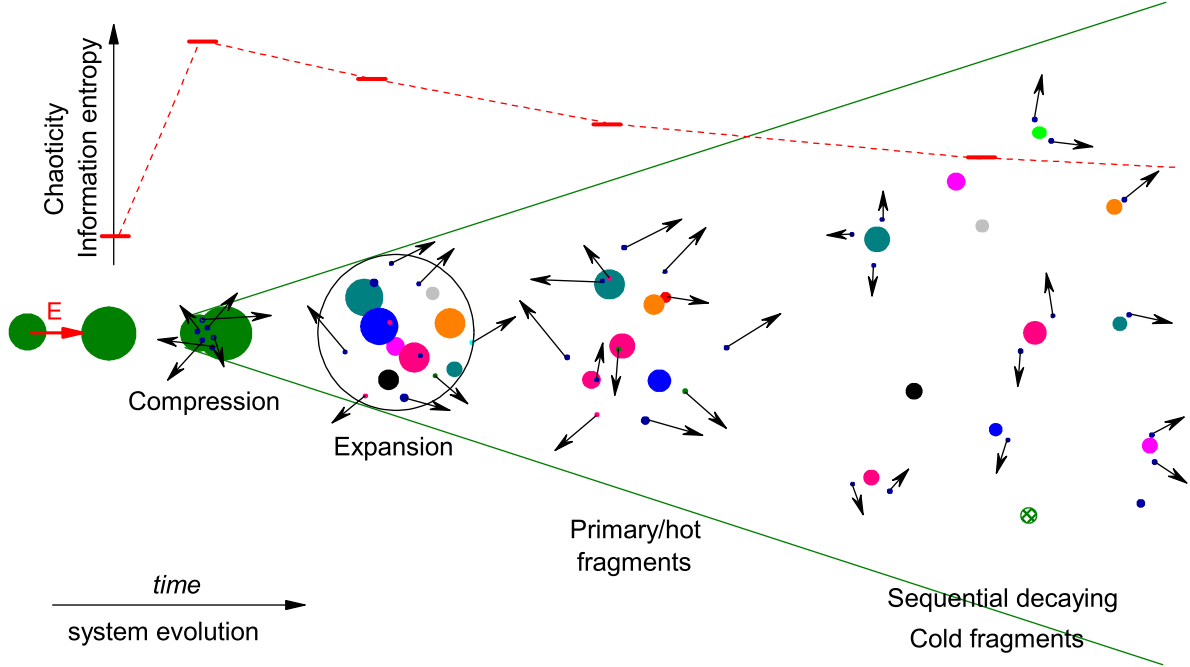,scale=1}
\end{minipage}
\begin{minipage}[t]{16.5 cm}
\caption{(Color online) Schematic drawing for time evolution of nuclear matter in heavy-ion collision, and chaoticity change in the process (red bars connected by dashed line). The chaoticity change (upper part in short bars connected by dashed line) is drawn according to the information entropy evolution in Ref. \cite{YG99PRL}. \label{ShmeetHic}}
\end{minipage}
\end{center}
\end{figure}

These features of Shannon information entropy theory make it favorite in the study of nuclear reaction, with the characteristics of dynamical evolution of nuclear matter. Similar to the diverse bunch of technologies of information communications, different kinds of reaction theories have been developed to describe the evolution of nuclear colliding systems with different aspects being particularly focused on \cite{IBD135}. In general, the processes of heavy-ion collisions are divided into two or three stages in different theories, including the collision and primary fragments formation, and the following de-excitation (see Figure \ref{ShmeetHic}). Considering the projectile and target nuclei firstly in the liquid drops, the collisions result in multifragmentation, which makes the chaoticity increases. The multiplicity information entropy was first defined by Y.G. Ma \cite{YG99PRL} and was applied to study nuclear reaction, which shows that the information entropy (or chaoticity) of system shows a trend of firstly increase and then decrease along the colliding time, where a liquid-gas phase transition happens. As a useful tool to measure uncertainty in a random variable which qualifies the expected information contained in a message, a constructive criterion will be provided by the Shannon information entropy for setting up probability distributions on the basis of partial knowledge \cite{CESh2}. It also suggests new observables and provides new methods to study the system evolution in heavy-ion collision by avoiding the physical limitation on system state (whether in an equilibrium state).

In this review, we discuss the applications of Shannon information entropy in heavy-ion collisions. In Section \ref{sec:BasicConcpt} the concepts of Shannon information entropy are introduced, and the applications of information entropy are presented. In Section \ref{sec:theories} the theoretical models for heavy-ion collisions are briefly introduced to aid the understanding of the theoretical descriptions of heavy-ion collisions, and in Section \ref{sec:SHEapp} the applications of Shannon information entropy are reviewed, and some prospects are outlined. A summary of this review is presented in Section \ref{sec:summary}.

\section{Shannon Information Entropy \label{sec:BasicConcpt}}

\subsection{Basic Ideas}
\subsubsection{Shannon Information Entropy}

It is a fundamental problem in information communication that the message selected in one point is either exactly or approximately reproduced in another point. The actual message is one selected from a set of possible messages, which is a significant aspect in communication. A communication system should be designed to operate for each possible selection. For a system with a set of possible events, which are discrete $S=\{E_{1}, E_{2}, ..., E_{n}\}$, with the corresponding occurring probabilities $P=\{p_{1}, p_{2}, ... p_{n}\}$, the definition of Shannon information entropy for a specific event is simply to take the (natural) logarithm of its probability. For the message space $S$, the information entropy is defined by C.E. Shannon as \cite{CESh1},
\be
H(S)=-\sum_{i=1}^{n}p_{i}\ln p_{i}. \label{eq:defShInfDis}
\ee
It is required that $\sum_{i=1}^{n}p_{i}=1$. If $p_{i}$ are equal, $p_{i}=\frac{1}{n}$, then $H$ increases monotonically with $n$, and one has the form of Hartley entropy,
\be
H(S)=\ln N
\ee
If $p_{i}$ are different, they play as a role of weighting the corresponding event \footnote{In Equation (\ref{eq:defShInfDis}), besides the natural logarithm, the logarithm can also be adopted. For consistence in description, We adopt the natural logarithm in this review.}.

If the events are in continuous state, for which the probability distribution is $P(m)$, one has,
\be
H(S)=-\int P(m)\ln P(m) dm, \label{eq:defShInfcnt}
\ee
where $\int P(m) dm = 1$ is required.

\subsubsection{Connection/Difference between Thermodynamic and Information Entropy}

The thermodynamical entropy, established by L. Boltzmann and J. W. Gibbs in the 1870s, is in the theory of statistical mechanics in the form of,
\be
S=-k \sum_{i}p_{i}\log p_{i},
\ee
where the subindex $i$ denotes microstate and $p_{i}$ is its probability in an equilibrium ensemble. If the microstates are in equal probability (a microcanonical ensemble), it becomes to the entropy on Boltzmann's tombstone,
\be
S=k \log W,
\ee
where $W$ is the number of microstates.
The Shannon information entropy has the same form as the thermodynamical entropy if one takes the logarithm in the definition. A direct connection can be made between the information entropy and thermodynamics entropy. ``Gain in entropy always means loss of information, and nothing more'' \cite{AllenMaxwell}. Though the entropy in information theory shares the same form and name of statistical mechanics, it is generally accepted that they are not the same. The information entropy, which can be calculated for any probability distribution, is applicable in a general manner, while the thermodynamic entropy is specifically for the thermodynamic probabilities with the physical system being in an equilibrium state.

\subsubsection{Joint Information Entropy}

The medical images will be taken as an example to introduce the concept of joint information entropy. For multimodality images, it can be assumed that the regions of similar tissue in one image corresponds to regions in the other image consisting of similar information expressed in such as gray values \cite{Woods1,Woods2}. The ratio of gray values in the two images will vary very little in a certain region, and the average variance of the ratio for all regions is minimized to achieve registration. In a {\it feature space}, an adaptive method was proposed by Hill \textit{et al} to plot the two-dimensional (2D) figure combining the gray values for all corresponding points in each of two images \cite{Hill}. In this method, in feature space the regions are defined based on the clustering one for registered images. A joint histogram of two images, from their gray values by dividing each entry in the histogram by the total number of entries, can be used to estimate a joint probability distribution \cite{MedInfo}. The definition of Shannon entropy for a joint distribution is  \footnote{In Ref. \cite{MedInfo}, they took the logarithm of the probability. To express the definitions in a uniform manner, the natural logarithm is used in this review.},
\be
-\sum_{i, j}p(i, j)\ln p(i, j),
\ee
where $i, j$ denote the numbers of regions of the horizontal and vertical directions in the 2D figure.
One can find that the transformation minimizes their joint entropy and registers the images.

\begin{figure}[htbp]
\begin{center}
\begin{minipage}[t]{11 cm}
\epsfig{file=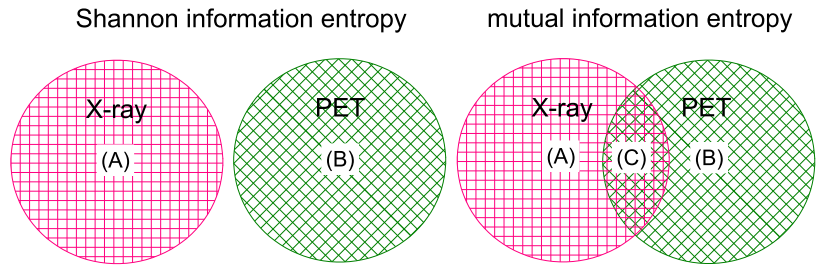,scale=0.9}
\end{minipage}
\begin{minipage}[t]{16.5 cm}
\caption{(Color online) Schematic plots for Shannon information entropy and mutual information entropy. Shannon information for an object obtained by X-ray image (A) and by PET image (B). The overlapping area (C) denotes the mutual information entropy. \label{MuEntpy}}
\end{minipage}
\end{center}
\end{figure}

\subsubsection{Mutual Information Entropy}

The mutual information entropy appears once entropy is introduced for handling complex radiometric relationships between images \cite{Collignon1,Viola,Wells}. Mutual information showed great promise and it became the most investigated measure for medical image registration. To introduce the mutual information entropy, firstly we take two pictures of one object by different techniques, for example the $X$-ray camera (labelled as $A$) and the positron emission computed tomography (PET) image (labelled as $B$) as shown in Figure \ref{MuEntpy}. With the $X$-ray image, the elements in the object can be identified. With the PET image, the structure defects can be found. The Shannon entropy for images $A$ and $B$ can be computed based on the probability distribution of gray value, labelled as $H(A)$ and $H(B)$, respectively. If the information in both $A$ and $B$ are combined, more information can be expected. For example, at a specific position, the defect of structure originates from the loss of which element. The best form to define the mutual information [$I(A, B)$] for the two images is,
\be
I(A, B) = H(B) - H(B|A). \label{MutDef}
\ee
$H(B|A)$ is the conditional entropy according to the conditional probabilities $p(b|a)$, i.e., the chance for gray value $b$ in image $B$ given that the corresponding voxel in $A$ has gray value $a$. If one interpretes entropy as a measure of uncertainty, $I(A, B)$ can be explained as ``the amount of uncertainty about image $B$ minus the uncertainty about $B$ about $A$ is known.'' One also can say that the mutual information is the amount by which the uncertainty about $B$ decreases when $A$ is given that the amount of information $A$ contains about $B$ \cite{MedInfo}. Since $A$ and $B$ are interchangeable, $I(A, B)$ is also the amount of information $B$ contains about $A$. That is the reason why $I(A, B)$ is called as {\it mutual information}. Besides the definition of mutual information in Equation (\ref{MutDef}), two more equivalent definitions can be used \cite{MedInfo}.

The second definition is \cite{MedInfo},
\be
I(A, B) = H(A) + H(B) - H(A, B). \label{MutDefE1}
\ee
The term $-H(A, B)$ means that maximizing mutual information is related to the minimization of joint entropy. The entropies of separate images are included in mutual information, which is an advantage over joint entropy.

The third definition is related to the Kullback--Leibler distance for two distributions $p$ and $q$ defined as $\sum_{i}p(i)\mbox{log}[p(i)/q(i)]$, which measures the distance between two distributions \cite{MedInfo}. The mutual information is defined as,
\be
I(A, B) = \sum_{a, b}p(a, b)\mbox{log}\frac{p(a,b)}{p(a)p(b)}. \label{MutDefE2}
\ee
In this manner of definition, the mutual information measures the distance between the joint distribution of the images' $p(a,b)$ and the joint distribution in case of independence of the images $p(a)p(b)$. It measures the dependence between two images by assuming that there is maximal dependence between the images when they are correctly aligned.

The mutual information has following properties \cite{MutProp}
\begin{enumerate}
  \item $I(A, B) = I(B, A)$,
  \item $I(A, A) = H(A)$,
  \item $I(A, B) \leq H(A)$, and $I(A, B) \leq H(B)$,
  \item $I(A, B) \geq 0$,
  \item $I(A, B) = 0$ only if $A$ is independent of $B$. It means that no knowledge is gained when one image is not in any way related to the other one.
\end{enumerate}

\subsubsection{Configurational Entropy}\label{sec:CE}
The configurational entropy was constructed in the functional space by M. Gleiser and N. Stamatopoulos \cite{CEDef12PLB}, which has been applied to nonlinear scalar field models in the feature of solutions with spatially-localized energy.

The construction of configurational entropy will be introduced in brief. For interests in structures with spatially-localized energy, considering a set of functions $f(x)\in L^{2}(\mathbf{R})$ and their Fourier transforms $F(k)$, the Plancherel's theorem is obeyed \cite{Plantheorem},
\be
\int_{-\infty}^{\infty} |f(x)|^{2}dx=\int_{-\infty}^{\infty} |F(k)|^{2}dk.
\ee
$f(x)$ should be square-integrable bounded. The model fraction $f(k)$ can be defined,
\be
f(k)=\frac{|F(k)|^{2}}{|F(k)|^{2}d^{d}k}.
\ee
The integration is over all $k$ where $F(k)$ is defined, and $d$ is the number of spatial dimensions. The model fraction $f(k)$ measures the relative weight of a given mode $k$.
For periodic functions, a Fourier series can be defined as $f(k)\rightarrow f_{n}=|A_{n}|^2/\sum|A_{n}|^2$, with $A_{n}$ being the coefficient of the $n$-th Fourier mode. The configurational entropy $S_{c}[f]$ is defined as,
\be
S_{C}[f]=-\sum f_{n} \ln (f_{n}),
\ee
which corresponds to the definition of Shannon information entropy $S=-\sum p_{i}\ln p_{i}$. The configurational entropy thus provides an informational content of configurations compatible with the particular constraints of a given physical system. If each mode $k$ carries the same weight, $f_{n}=1/N$ and the discrete configuration entropy reaches the maximum at $S_c= \ln N$. If only one mode $S_{c} =$ 0.

The continuous configurational entropy can also be defined. For general, for non-periodic functions in an interval $(a, b)$ it will be,
\be
S_{C}[f]=-\int \tilde{f}(k)\ln [\tilde{f}(k)]d^{d}k,
\ee
in which $\tilde{f}(k)=f(k)/f(k)_{\mathbf{max}}$ and $f(k)_{max}$ is the maximum fraction. The normalized function $\tilde{f}(k)$ guarantees that $\tilde{f}(k)\leq$ 1 for all mode $k$. The integrand $\tilde{f}(k) \ln \tilde{f}(k)$ is called as the configurational entropy density. The form of definition for $S_{C}[f]$ is similar to the Gibbs entropy, which is defined of nonequilibrium thermodynamics for a statistical ensemble with microstates with probability $p_{i}=\exp(-E_{i}/k_{b}T)/\sum\exp(-E_{i}/k_{b}T)$ (${i}$ is the energy of the $i-$th microstate and $k_{b}$ is Boltzmann's constant) \cite{GibbsEntrpy}.

\subsection{General Applications of Information Entropy \label{sec:GenIntr}}
The conciseness of the information entropy based analysis methods meets the requirement where the system is complex and the traditional theories have difficulties. Among the various applications of the entropy information theory, as have been briefly reviewed in a recent work \cite{entApp17}, some applications to interesting topics, which are divided into the general sciences and physics, are going to be briefly introduced in this section.
\subsubsection{Applications in General Sciences}
For simplification, we itemize the selected topics in this section, which include the medical iconography, the species geographic distributions, the emergences in artificial society, the machine learning, the molecular docking, and the classification of ground water, etc.
\begin{itemize}
\item \textbf{Medical Iconography}. In medical iconography, the registration of images from different modalities such as computed tomography (CT), PET, and magnetic resonance (MR) will comprehensively capture their own unique information (mass/bone density, metabolic rate, proton density, respectively) about the same patient, and it is expected to help diagnosis, staging, prognosis, and to detect the ranges of organ motion and changes in tumor position, size, and shape \cite{JF01,JF02}. Registration algorithms are central to these processes. In a pixel-based approach, registration takes advantage of the gray-level information of all pixels of these images based on mutual information, and registration accuracy depends on several factors such as pixel size and field-of-view \cite{JF03,JF04}. In all the modalities for registration, joint information entropy analysis is used to deal with the classification of gray value of the image in the feature space, which measures the disperse of probability distribution. If one combines the images of two or more kinds of tomography imaging, for example MR (A) and CT (B), the mutual information can give more precise information, i.e., in (A) the information containing (B) can be reflected.

    The advantage of mutual information over the joint information is that the mutual information includes the entropies of the separate images. For the overlapping parts of the images, mutual information and joint entropy are both computed, and the measures are therefore sensitive to the size and the contents of overlap. Pluim \textit{et al} reviewed the mutual-information-based registration of medical images in aspects of the methodology and of the application \cite{MedInfo}.

\item \textbf{Species Geographic Distributions.} For some species, the detailed presence/absence occurrence data allow the use of a variety of standard statistical techniques to predictively model the species environmental requirements and geographic distributions, while for most species the absence of data hinder the researchers to find the environmental requirements and geographic distributions for them. Based on the maximum entropy (Maxent) method, a machine learning method is proposed to model species geographic distributions with presence-only data \cite{MaxtentInfo}, and becomes a general-purpose method to predict or infer from incomplete information. The origin of Maxent lies in statistical mechanics, and is active in areas of exploration applications in diverse areas such as astronomy, image reconstruction, portfolio optimization, signal processing, and statistical physics. Employing a simple and precise mathematical formulation, Maxent has a number of aspects which make it well-suited for species distribution modeling. The Maxent modeling approach succeeds in many applications with presence-only datasets, and attracts further research and development (see a recent review \cite{Maxtent-1}).

\item \textbf{Emergences in Artificial Society}. An emergence is a general phenomenon with an important characteristics of complex dynamic systems \cite{Emerg}. It is always connected with the concept of self-organization, which may be due to the local nonlinear self-organized interaction between individuals in a complex dynamic system \cite{Prok}. Or one can say that the global properties, behaviors, structures or patterns of a complex system arise from localized individual behaviors. In real society, an emergency represents an event or situation that in potential can pose immediate risks to health, life, property or the environment \cite{EmerHN}, which may give rise to other emergences and have a negative influences on the individuals and even can result in a great loss of life and property, e.g., an outbreak of H7N9, H1N1 or SARS. Emergencies like H7N9 and SARS are of nonrepeatability and high-cost of experiments in the real society, which makes people to naturally borrow the idea of artificial society and parallel societies to aid emergency management \cite{InteSoc}. Emergences in general can be sorted to the weak emergences and strong emergences. A strong emergence is a macro-level property, which in principle cannot be identified from the sum of micro-level properties and it's hard to measure such a strong emergence. A classic case of strong emergence is the emergence of consciousness such as the qualia of pain from the neurobiological processes \cite{MacroSim}. The weak emergences are in principle identifiable from micro-level individuals \cite{MacroMicro} and are derivable. The micro-to-macro derivable tracks of weak emergences should be non-trivial, which can be amenable to computer simulations \cite{MicroSim}. Tang and Mao \cite{ArtiInfo} proposed a metrics based on information entropy to measure the emergences in an artificial society model (the zombie-city), which overcomes the diversity of artificial societies including the emergences of attribute, behavior and structure. Three metrics were proposed to denote the three kinds of emergences (attribution/behavior/structure) in artificial societies, and to quantitatively measure emergences.

\item \textbf{Label Ranking in Machine Learning.} In machine learning, the Label Ranking (LR) problems are becoming increasingly important. It focus on the problem of learning a mapping from instances to rankings over a finite number of predefined labels, for example, the typical ranking of a set of restaurants or parking according to the preference of a given person. Compared to the traditional classifying problems \cite{ClassProb}, LR is interested in assigning a complete preference order of the labels to every example. In machine learning tasks, it is an essential task to develop accurate LR models for a proper data preparation. Some algorithms are unable to deal with numeric variables, which requires a discretization of the numeric variables beforehand. Discretization means the partition of a given interval into a set of discrete sub-intervals, and split continuous intervals into two or more sub-intervals which can be treated as nominal values. A good discretization method should well consider the balance between the information loss and partition numbers \cite{GoodDiscr}. Depending on whether or not they involve target variable information, discretization methods are typically organized in two groups, i.e., the supervised and unsupervised discretization, respectively. S\'{a} \textit{et al} proposed a pre-processing method of a  supervised discretization for ranking data based on the minimum description length principle (MDLP) \cite{MDLP} and the extended version of Minimum Description Length Principle (MDLP-R) \cite{EDiRa}. Adopting the concept of entropy, the MDLP was also extended to the version of Entropy-based Discretization for Ranking (EDiRa) \cite{EDiRa,DataRankSa}. The comparison between the results shows that EDiRa behaves better in many scenarios and is also more robust \cite{DataRankSa}. Sang \textit{et al} have also introduced the entropy based method in dealing with the discretization of high-dimensional data \cite{highDim}, which can achieve higher classification accuracy and yields a more concise knowledge of the data especially for high-dimensional datasets than existing discretization methods.

\item \textbf{Molecular Docking.} Molecular docking predicts the conformation of a ligand within the active site of a receptor and searches for the low-energy binding modes \cite{DefDocking}, which is widely used in virtual screen. The docking model has received wide interests and a lot of scoring functions have been proposed \cite{ScoFunct}. As the core of molecular docking, the scoring function helps a docking program to efficiently explore the binding space of a ligand, and optimize the process of finding the best position of a ligand in the binding site of a receptor. It also evaluates the binding affinity once the correct binding pose is identified \cite{RolFun}. The docking program and scoring functions are compared \cite{DefDocking}, while they are found to cannot generally applicable for all the situations due to the very complicated interactions between ligands and receptors. Meanwhile, the docking models should be simplified for an acceptable computing time. Li \textit{et al} have proposed an iteration scheme in conjunction with the multi-population evolution, in which the entropy-based method is adopted as the searching technique to control the narrowing of searching space for each population in molecular docking \cite{DefDocking}. When the searching space in the best population is reduced to a given tolerated area, the global optimal solution can be obtained approximately. The method improves the docking accuracy over the individual force-field scoring greatly. Comparing with other softwares, the best average root-mean-square deviation can be reached within a good average computing time.

\item \textbf{Classification of Water.} It makes the classification of water difficult since various methods exist for drinking water quality criteria and decision making. The complex physicochemical parameters including toxic heavy metals like As, Pb, Cr, Ni, non-toxic elements Fe, Zn, Cu, Mn, cations Na, K, Ca, Mg, Ba, and anions F, Cl, SO$_4$, NO$_3$, pH, EC, and TDS, etc., are usually adopted to calculate the drinking quality rank of water samples. The most popular and commonly used methods, water quality index (WQI), integrate water quality variables adopting the various water quality indices. These methods are complained to have a few drawbacks since the WQI is dramatically influenced by some parameters without valid scientific justification. The limitation of the fixed weight assigned to different parameters maybe result in wrong decision. It was proposed that the weighting should be varied basing on season, ambient temperature, rainfall and water intake of individual, residential, occupational and other environmental factors, which requires an advanced classification method to account for imprecise and vague information in decision making on drinking water quality. The information entropy is introduced to classification of groundwater by Kamrani \textit{et al} to measure the amount of useful information with sample \cite{Water}. In their entropy based WQI, the entropy is used to determine the weights of each indicator based on the evaluation of the variation degree of every evaluation indicator value. It makes the entropy coefficient method be an objective empowering method.
\end{itemize}

\subsubsection{Applications in Physics}
The configurational entropy makes a connection between the dynamical and informational contents of physical models with localized energy configurations \cite{CEDef12PLB}, which is useful in the questions of the solitons and bounces in one spatial dimension, and critical bubbles in three spatial dimensions, typically the first-order phase transitions can be solved. Many fields of physics, like high energy physics, cosmology, condensed matter physics, etc, are suggested to adopt this configurational entropy measurements. The configurational entropy has been adopted in the study of non-linear physical system \cite{Soli06,Soli07,Soli08,Soli09}, to distinguish configurations with energy-degenerate
spatial profiles \cite{CEAPP01}, to study compact astrophysical systems \cite{Soli06,Soli10,Soli11}, the scalar glueballs \cite{Soli04}, AdS-Schwarzschild black holes \cite{Soli12,CEAPP03}, the phase transition of field configurations from single-kink solution to double-kink (multi-kink) solution \cite{CEAPP02}, and to determine the Higgs boson mass \cite{Soli16} and the axion mass in an effective theory at low energy regimes \cite{Soli17}.

Some examples will be introduced in brief. The interested readers can refer to the original materials cited. Silva and Rocha's work \cite{Soli01} deals with a special class of solitons, known as the Korteweg-de Vries (KdV) solitons, which are spatially localized pulses, of finite energy, propagate with at most tiny distortions of their shape. The analysis shows that, even for the same solitonic solution of the KdV equation, the variations in the spatial shape induced by choices of relevant parameter are non-trivial in the informational sense. The amount of information needed to describe the system is at the most compressed state, but neither the minimum width nor the maximum one. More works used the configurational entropy method to demonstrate a high organisational degree in the structure of the configuration of the system, and to the refinement of topological defects by selecting the most appropriate parameters corresponding to the most organized system \cite{CEInfor1,CEInfor2,CEInfor3}. In the cold quark-gluon plasma (cQGP) system within a mean-field quark-gluon plasma model, it is also shown that the larger the energy of the condensate, the smaller the width of the informational optimal soliton is \cite{Soli01}. The configurational entropy indicates that the informational content of the soliton spatial profile is more compressed. Braga and Rocha \cite{CEAPP03} also showed the evidence based on configurational entropy that the smaller the black holes, the more unstable they are and furnishes a reliable measure for stability of black holes. In a recent work, Braga and Rocha \cite{AdSQCDdual18} paved a way into the bottomonium and charmonium (quarkonia) production from the point of view of the configurational entropy in the AdS/QCD correspondence, where the configurational entropy has provided data regarding the relative dominance and the abundance of the bottomonium and charmonium states whose underlying information is more compressed in the Shannon's theory meaning. And the results from the configurational information entropy also identify the lower phenomenological prevalence of higher S-wave resonances and higher masses quarkonia in Nature. Karapetyan \cite{KaraCEAPP04EPL,KaraEPL2017} studied the configurational entropy dependence on impact parameter in the particle production in quark-antiquark scattering in a simple dipole color-glass condensate ($b-$CGC) model (one can refer to the applications of $b-$CGC model in deep inelastic scattering of lepton induced reactions and diffractive process \cite{cgc-scat} and proton-nucleus collisions at RHIC and LHC \cite{cgc-pA}). It is illustrated the nuclear configurational entropy is a way to indicate the onset of the quantum regime, and in potential can be used to study the encompasses quantum mechanics fluctuations.

\section{Theoretical Models for Heavy-Ion Collisions \label{sec:theories}}

A finite nucleus can be viewed as a charged liquid drop with the composition being different ratios of protons and neutrons. In heavy-ion collisions, the impact between the projectile and target nuclei destroy the order of protons and neutrons, which increases the chaoticity of system. In this process, the nuclear liquid may be vaporized to a gas state. Like the heated water, the phenomenon of liquid-gas transition may happen (see Figure \ref{ShmeetHic}). Theoretically, the processes of heavy-ion collisions are usually divided into two or three stages. A ``fireball'' is always used to describe the zone of colliding. In the range of medium-energy regime, i.e., from several tens to several hundreds MeV/u, both light particles and intermediate-mass fragments (IMFs, $Z > 2$) are produced. At the relatively low energies, the IMFs may be produced at mid-rapidity, while copious clusters will be produced in central higher-energy collisions during the system expansion, and highly excited projectile-like fragments will be produced in the peripheral collisions. Typically, the IMFs carry a major part ($\sim$ 50\%) of nucleons involved \cite{ThM01}, which is only a partial information of the system. If the fragments from the light particles to IMFs as the events in the colliding system, and their cross sections (which are measurable in experiments) as the probabilities, the information entropy analysis can be naturally constructed and applied in heavy-ion collisions.

To study the nuclear matter changes in heavy-ion collisions, theoretical simulations take important roles in extracting information from the changing situations of the colliding system. Theoretical simulations are also important in determination of key processes and basic parameters, which cannot be measured directly in experiments. Before the descriptions of applications Shannon information entropy in heavy-ion collisions, we briefly introduce the generally used theoretical models describing the processes of heavy-ion collisions, which include the dynamics models, the thermaldynamics models, the statistical models, and the models for the decay process.

\subsection{Dynamics Models \label{sec:Mtran}}

The microscopic dynamical models is needed to describe fragments production since they are formed in dynamical reactions, and the equilibrium of the system is not guaranteed. The quantal nature of the many-body nuclear system makes it difficult to develop a microscopic dynamical model without simplifications. The dynamical models are characterized by the description of the nucleons transportation in nuclear field governed by interactions, which is also called as the transport models. The present versions of transport models consider different characteristics related to the most important aspects of heavy-ion collisions, and can only seek reasonable successes in specific cases. The transport models are important in extracting physics information from experiments. Many different transport codes emerge, in which different physical influences have been considered and improved in them. In transport models, the equation of state (EoS) of nuclear matter is taken as one of many key inputs to model the heavy-ion collisions. Within the incident energy range from the Fermi energy to relativistic energy, transport theories serve as important roles in obtaining valuable information from heavy-ion collisions. In this review, we concentrate on the understanding of nuclear fragmentation reactions.

To describe nuclear fragmentation reactions, it is generally required that the following should be included, i.e., the framework of the time evolution, the dynamical bifurcations, the basic quantum statistics, the macroscopic nuclear properties, the thermal nuclear properties, the interactions, the particle emission from hot nuclei, and the correlations and fragmentation mechanisms \cite{ThM01}. The features of transport models are as follows,
\begin{itemize}
\item A mean-field picture is adopted in most of the transport models, exemplified by the time-dependent Hartree-Fock (TDHF) treatment or the semiclassical analogue via the Vlasov equation. The models should include both the single-particle motion in a mean field and Pauli-suppressed two-nucleon collisions, since the direct two-nucleon collisions become more important with incident energy around the Fermi-energy.
\item Beyond the mean-field description, the many-body correlations are desired to be included in dynamical models, in particular for light fragments like the $\alpha$ particle. It is one of the advantage of molecular dynamics since the correlations are included automatically.
\item The dynamical bifurcations including spontaneous fluctuations and the associated trajectory branchings should be treated for fragmentation. Most of the transport models employ some stochastic agent to avoid the difficulties of the standard coupled-channel treatment.
\item The basic quantum statistical features of Pauli blocking and Fermi motion of nucleons should be incorporated.
\item The most important macroscopic nuclear properties, usually the binding energy and density distributions (or radius), are reproduced.
\item The statistical features play a significant role in the determination of the relative fragment yields. The nuclear level density is important in dynamical models.
\item The models contain only a minimal number of parameters. It is important for the dynamical model that the interaction employed can yield a liquid-gas phase transition in uniform matter. For some dynamical models, it is an important feature that only the interaction govern the reactions.
\item The de-excitation of hot fragments, which emit light-particles over a time scale much longer than the violent collision, is required in dynamical models. The decay process requires a de-excitation treatment to each of hot fragments formed in the collision stage.
\end{itemize}

The nowadays dynamics theories start from early works to describe the heavy-ion collisions \cite{trans31,trans32,trans33}, including the mean-field models with fluctuations and the models of molecular dynamics. Two families of transport methods are formed: (1) The evolution of one-body phase-space density is formulated by the Boltzmann-Vlasove type formulates under the influence of mean field, which is usually referred as the Boltzmann-\"{U}ehling-Uhlenbeck (BUU) theories. (2) The other type uses the molecular dynamics, in which the coordinates and momenta of nucleons are controlled under the action of a many-body collision term, which is referred as the quantum molecular dynamics (QMD). The complexity of the BUU and QMD theories makes the corresponding simulations involve many choices and strategies. Basically, there are two main ingredients in transport calculations. The first one is the mean field which is related to the EoS, and the second one is the nucleon-nucleon (NN) collisions which have different influences on the reaction dynamics. In spite of the remarkable successful applications of transport models, the many versions of transport models have given different predictions for physical observables with similar nuclear input. In this review, we briefly introduce the models. For details of the approaches, the readers are referred to more original references. A recent comparison of results from various transport models with same inputs also help to understand the physics for heavy-ion collisions \cite{trans34}.

\subsubsection{The BUU Model}

The time-dependent Hartree-Fock (TDHF) treatment, which is in the microscopic approaches \cite{TDHF1,TDHF2,TDHF3,TDHF4,TDHF5,TDHF6,TDHF7,TDHF8}, provides a useful starting point for nuclear dynamics. The BUU theory includes the residue interaction, in which the collisionless mean-field evolution is argumented by a Pauli-blocked Boltzmann collision term. The BUU theory can be derived from  Born-Bogoliubov-Green-Kirkwood-Yvon, or more effectively derived from the real-time Green's-function Martin-Schwinger formalism to reach the Kadanoff-Baym equations \cite{trans35,trans36,trans37}. The evolution of the one-body phase-space density $f(\vec{r}, \vec{p}; t)$ obeys the BUU equation,
\be
(\frac{\partial}{\partial t}+\frac{\vec{p}}{m}\cdot\nabla_{r}-\nabla_{r} U \cdot\nabla_{p})f(\vec{r}, \vec{p}; t)=I_{coll}[f; \sigma_{12}]. \label{eq:BUU-phasespace}
\ee
The mean field $U[f]$ is usually formulated as a density functional which depends on the baryon density, the spin and possible isospin. $I_{coll}$ is the two-body collision term,
\be
I_{coll}=\frac{1}{(2\pi)^6}\int dp_{2}dp_{3}d\Omega|v-v_{2}|
\frac{d\sigma_{12}^{med}}{d\Omega}
(2\pi)^3\times\delta(p+p_2-p_3-p_4)
\times[f_3f_4(1-f)(1-f_2)-ff_2(1-f_3)(1-f_4)],
\label{eq:buuIcoll}
\ee
in which $\sigma^{med}$ is the in-medium nucleon-nucleon cross section. The UU (Uehling-Uhlenbech) in the BUU equation describes the Pauli-blocking factors in the gain and loss terms of collisions, which is also adopted in the Boltzmann-Nordheim-Vlasov (BNV), Vlasov-Uehling-Uhlenbeck (VUU), and Landau-Vlasov (LV) models \cite{trans34}. The distribution function is resolved in the terms of a (large) number of discrete test-particles as,
\be
f(\vec{r}, \vec{p}; t) = \frac{1}{N_{TP}}\sum_{i=1}^{N_{TP}A}
g[\vec{r}-\vec{r}_{i}(t)]\tilde{g}[\vec{p}-\vec{p}_{i}(t)], \label{eq:buudensityf}
\ee
where $A$ is the number of nucleons, $N_{TP}$ is the test particle number per nucleon; $\vec{r}_{i}$ and $\vec{p}_{i}$ denote the coordinates and the momenta of the test particles, respectively; $g$ and $\tilde{g}$ are the shapes of the coordinate and momenta. The coordinate is often taken as $\delta$ functions (as point particles), while the Gaussian or triangular shapes are also used to reduce the number of test particles, as well as make the distribution smoother. Approximatively, the test particle obeys the following Hamiltonian equations of motion under the influence of mean field,
\be
d\vec{r}_i/dt=\nabla_{\vec{p}_i}H;
~~~~~~
d\vec{p}_i/dt=-\nabla_{\vec{r}_i}H. \label{eq:buuHamilton}
\ee

The collision term is commonly simulated randomly, with the test particle collisions at $\sigma^{'} = \sigma^{med}/N_{TP}$. The test-particle configuration is sampled to find whether the test particles are close enough to make a collision or not. For each ``attempted'' collision, the phase-space occupation for final states is calculated for the Pauli blocking, with the Pauli-blocking probability being calculated in most cases from $1-(1-f_3)(1-f_4)$. A recent review for BUU model can be found in Ref. \cite{RrpGiBUU}, in particular for the Giessen BUU (GiBUU) model.

\subsubsection{The Molecular Dynamics Models}

The molecular-dynamics many-body methods provide a more direct connection to the observable. According to its characteristics, the frequently used molecular dynamics can be divided into the classical molecular dynamics (CMD), the quasi-classical molecular dynamics, the QMD, the constrained molecular dynamics (CoMD), the Fermionic molecular dynamics (FMD), the antisymmetrized molecular dynamics (AMD), and quantal langevin dynamics, etc. Reviews of these models can be found in many articles, such as Refs. \cite{ThM01,trans34}

{\it \textbf{Classical Molecular Dynamics}.}
The CMD solves the classical equation of motion for a system of $A$ particles,
\be
\frac{d}{dt}\vec{r}_{i}=\{\vec{r}_{i}, H\},~~~~\frac{d}{dt}\vec{p}_{i}=\{\vec{r}_{i}, H\}.
\ee
The many-body Hamiltonian is,
\be
H\{\vec{r}_{n}, \vec{p}_{i}\}=\sum_{i=1}^{A}\frac{\vec{p}_{i}^{2}}{2m_{i}}+\sum_{i<j}V(|\vec{r}_{i}-\vec{r}_{j}|),
\ee
where $V(r)$ is the nucleon-nucleon potential, in which a short-range repulsive part and a long-range attractive part is usually consisted. The CMD model retains all the orders of many-body correlations in the classical level, which provides a useful insight into the general features of fragmenting finite systems. While the CMD model cannot include the fermion nature of nucleons. A so-called Pauli-potential is introduced to emulate the exclusion principle, which makes the CMD model be the quasi-classical molecular dynamics.

{\it \textbf{Quantum Molecular Dynamics}.}
The QMD model goes beyond the deterministic molecular dynamics, in which a Pauli-blocked collision term is introduced in a manner similar to the BUU-type treatments. The QMD approach is related to CMD, for which the two- or many-body interactions between nucleons are formulated in Hamiltonian. The nucleons are assumed as particles with finite widths usually in the shape of a Gaussian function. The QMD model can be derived from a time-dependent Hartree (TDH) theory, in which a trial wave function $\phi(\vec{r}; t)$ and positions [$\vec{R}_{i}(t)$] and momenta [$\vec{P}_{i}(t)$] are taken as variational parameters \cite{trans38},
\be
\Psi(\vec{r}_1,...\vec{r}_N; t) = \prod\phi_{i}(\vec{r}; t),  \label{eq:qmdWaveFunc}
\ee
where $\phi_{i}(\vec{r}; t)$ is,
\be
\phi_{i}(\vec{r}; t)=\frac{1}{(2\pi)^{3/4}(\Delta x)^{3/2}}
\mbox{exp}\{-\frac{[\vec{r}_i-
\vec{R}_{t}(t)]^2}{2(\Delta x)^2} + i\vec{r}_{i}\cdot\vec{P}_{i}(t)\}.\label{eq:qmdWaveFunctest}
\ee
The variation of wave function $\Psi$ leads to the equations of motion for the nucleons as the same as Equation (\ref{eq:buuHamilton}). Large differences in the propagation are not expected for one-body observables. For the antisymmetrized molecular dynamics model \cite{trans39} or the fermionic molecular dynamics model \cite{trans40}, the antisymmetrization of the wave packets is considered as derived from TDHF with a Slater determinant of Gaussians as a trial wave function. The equation of motion becomes more complicated, in which a norm matrix is involved because the wave-packet overlap changes when they move \cite{trans39}. The constrained molecular dynamics model takes an effective interaction of wave packet \cite{trans41}.

The two-body collisions in the QMD approaches are simulated in a similar way in BUU. Nucleons collide with the probability described by $\sigma^{med}$ in the intrinsically stochastic manner, and a collision affect the distribution function more considerably than the test particle collisions in BUU. The event-by-event fluctuations are induced by two-nucleon collisions in QMD, while is not in BUU. One run performed with a initial state in simulation, and is called as an ``event''. A large numbers of ``events'' with different initial states are averaged to obtain the mean and variation of the final result.

For the QMD-type models \cite{trans38,QMDtp1,QMDtp2,QMDtp3,QMDtp4,QMDtp5,QMDtp6,QMDtp7,QMDtp8,QMDtp9,QMDtp10}, the initial configurations of nuclei, which are obtained from initialization simulations, are selected when the similar binding energies and charge radii to the experimental data are obtained. Sometimes, a minimum distance between two arbitrary nucleons is required to give a more uniform initial phase-space distribution. In some models, a frictional cooling method is used to reach a ground-state initialization. More information about the initiation of different codes can be found in Ref. \cite{trans34}. The initial density distribution of nucleus is usually taken as the Woods-Saxon form,
\be
\rho(r)=\frac{\rho_{0}}{1+\mbox{exp} [(r-R)/a]}, \label{eq:WStypeDens}
\ee
where $R$ is the nuclear radius, $\rho_{0}$ is the saturation density, and $a$ is the diffuseness parameter.

{\it \textbf{Constrained Molecular Dynamics}.} The exact treatment of the Pauli principle requires significant computational power for large systems. The CoMD model provides an approximation implementation of the Pauli principle \cite{trans41,CodeCoMD}. A stochastic process is adopted to modify the usual QMD to avoid the violation of the Pauli principle, in case when the phase-space density $f_{i}$ around a nucleon $i$ is greater than 1. For the $i-$th nucleon, its momenta and that for other nucleon(s) are changed so that the Pauli principle $f_{i} \leq $ 1 is satisfied after several trials. A better description of the properties of hot nuclei may be achieved by CoMD than QMD. The charge distribution of intermediate-mass fragments are reasonably reproduced, except for the problems in the multi-particle multiplicities \cite{CoMD14PRC1,CoMD14PRC2,CoMD17PRC}.

{\it \textbf{Fermionic Molecular Dynamics}.}
The FMD model adopts a true quantum treatment for the many-body state \cite{FMD65,FMD66,FMD67,FMD68,FMD69}. In it an antisymmetrized Slater determinant of wave packets in a Gaussian form is taken,
\be
\varphi_{i}(\vec{r})\sim \mbox{exp}[-\nu_{i}(\vec{r}-\vec{Z}_{i})],
\ee
where $\{\vec{Z}_{i}\}$ and $\{\vec{\nu}_{i}\}$ are the wave packet centroinds and widths, for which the equation of motion can be derived from the time-dependent variational principle. The nucleons move in a mean field, for which the spin and isospin degrees of freedom may be included. The FMD wave function provides a reasonable approximate of ground-state nuclei with the energy of the constrained wave function being minimized \cite{FMD67}. The binding energy and radii of nucleus can be well reproduced with a reasonable effective interaction. The Pauli principle is perfectly incorporated in the FMD model by a Slater determinant \cite{FMD67}. While the deterministic character of FMD results in its drawback which does not offer a natural description of dynamical bifurcations. For example, nucleon emission occurs with some probability in case that the nucleon still remains in the source with the complementary probability \cite{ThM01}.

{\it \textbf{Antisymmetrized Molecular Dynamics}.} The AMD model \cite{AMD76,AMD77,AMD78} is similar to FMD. A  Slater determinant is used to represent the system and a part of the equation of motion is derived from the time-dependent variational principle. The difference between AMD and FMD is that stochastic terms have been added to the equation of motion in AMD, and many configurations appear through the reaction dynamics. In AMD, $\{\nu_{i}\}$ of the single-particle wave packets are taken as constant, which simplifies the computational time. It is guaranteed that there is no spurious coupling of the internal motion and the centre-of-mass motion of a cluster of a nucleus. With the extensions such as the parity and angular-momentum projections, a quite good description of both the basic properties but also many detailed structures can be achieved. The ground-state of nuclei, the radii, as well as the excitation level spectra of light nuclei can be described \cite{AMD79}.

The description of the two-nucleon collisions in AMD is similar to QMD. But the wave packet centroids $\{\vec{Z}_{i}\}$ do not represent the positions and momenta of nucleons. The physical coordinates are introduced as nonlinear functions of centroid  and the two-nucleon collisions are performed by them \cite{AMD77}. The Pauli-forbidden phase-space regions are not allowed as a final state of a collision, and the collision is canceled. Another typical difference between the AMD and QMD models is that the physical momentum in AMD is the momentum centroid of a Gaussian phase-space distribution, while the momentum variable in QMD represents the definite momentum of a nucleon.

The AMD model has been extended by Lin \textit{et al} to carefully treat the Fermi motion as a quantum fluctuation, in which a quantum branching of the wave packets is called as a diffusion process in the nucleon propagation in a mean field \cite{AMD80,AMDFM16PRC,AMDFM36}. The quantum fluctuation in each collision process is also taken into account. The extended version is called as the AMD-FM version \cite{AMDFM16PRC}. The feature of the AMD-FM version is that a new Fermi motion is assigned as a momentum fluctuation many times in the diffusion process and each collision throughout the calculation, which is quite different from that in most other transport models, in which the Fermi motion is added to nucleons only in the initial nuclei. The three-nucleon (3N) force is incorporated into the AMD-FM version by R. Wada to study the high energy protons, which suggests the first time that the 3N interaction is important in intermediate heavy-ion collisions in a full transport simulation \cite{AMDFM17Wada}.

The AMD model has been successfully applied to fragmentation reactions, such as central collisions in the energy region of several tens of MeV/nucleon for light and in-medium systems \cite{AMD80,AMDFM16PRC,AMDFM17Wada,AMD87,IBD144,Hipse-mocko,IBD163,IBD166}.
The huge time and CPU requirements for the simulation of a relatively larger system, for example the reactions of Au, hindered the  applications of the AMD model.

The comparison between the many BUU and QMD codes \cite{ThM01,ThM02,ThM03}, due to the different philosophies to deal with the fluctuations and correlations, shows that the predictions have a large spread. It is indicated that agreement of a simulation with an experimental observable alone may not serve as validation that the extracted physical parameters are reliable \cite{trans34}. It is also particularly important to understand and improve the predictions of transport simulations today.

\subsection{Thermaldynamics and Statistical Models\label{sec:TherStaMod}}
\subsubsection{Thermaldynamics Models \label{sec:MTher}}

In most general considerations of thermal dynamics models, three stages can be considered in the process of nuclear fragment production: (a) the formation of an intermediate highly excited nuclear system, (b) the expansion of the system and its disassembly into fragments, and (c) the sequential decay of hot primary fragments. The expansion time depends on the initial conditions, such as the incident energy, and the asymmetry of the system \cite{QMDtp10,CoMD14PRC2,BA98PRC,Xiao02PLB,Pili12PRC,Hud14EPJA,driftXZG17PRC}. It can be short in  case of central collisions of symmetric nuclei ($\sim$ 50 fm/c), and also long to few 100 fm/c in the case of peripheral heavy-ion collisions \cite{therm42}. Intensive exchange of mass, charge and energy between different parts of the system takes place during the expansion stage, for which partial thermodynamic equilibrium can be established prior to the break up. It will be reasonable to use a statistical representation to describe the final state ensemble of fragments at the excitation energy range of about 1 - 10 MeV/u. All the possible final states will be sorted out and their relative probabilities are calculated according to the corresponding statistical weight, for which the total energy, mass number and charge contained in the break-up volume should be considered for the break-up channels.

A hypothesis is adopted in the thermodynamic models that the system is in dilute media governed by short-range attractive forces. The formation of clusters are an result of the interparticle interactions \cite{FntMall15}. The most important part of the thermal dynamical models are the break-up stage, which describes the explosion of the equilibrated source. The canonical/macrocanonical/microcanonical ensemble theories have been adopted to describe this stage \cite{ENST01,ENST02,ENST03,ENST04,IBD145,IBD146}.

\begin{figure}[htbp]
\begin{center}
\begin{minipage}[t]{15 cm}
\epsfig{file=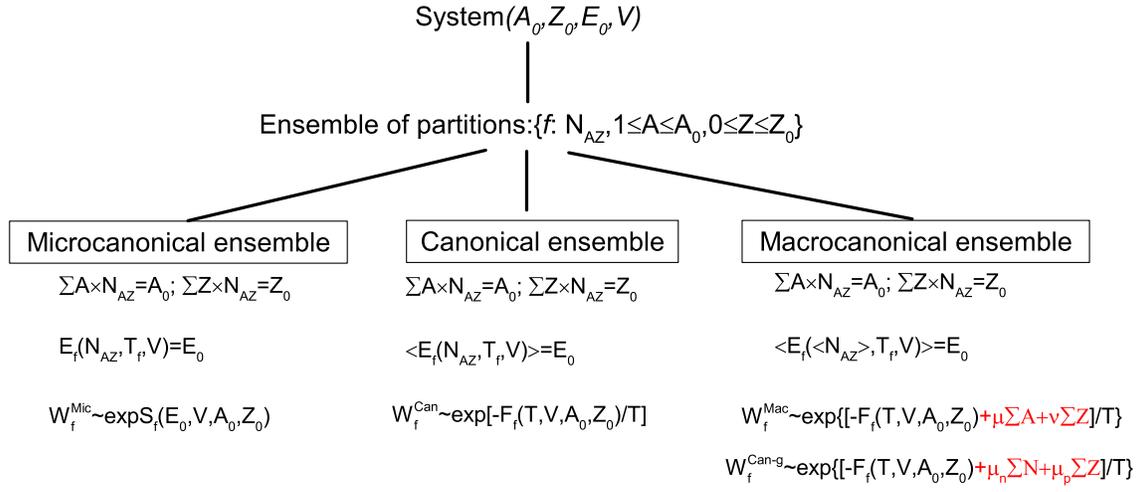,scale=0.8}
\end{minipage}
\begin{minipage}[t]{16.5 cm}
\caption{(Color online) Schematic classification of statistical ensembles for describing the break-up of a nuclear system ($A_0, Z_0, E_0, V$) with $A_0$, $Z_0$, $E_0$ and $V$ being the mass number, charge number, total energy and system volume, respectively. The $W^{Mic}_{f}$, $W^{Can}_{f}$, $W^{Mac}_{f}$, and $W^{Can-g}_{f}$ refer to the partition probability of microcanonical, canonical, macrocanonical, and canonical+grandcanonical limits ensembles, respectively. \label{canonical}}
\end{minipage}
\end{center}
\end{figure}

The description of canonical formulism can be found in Ref. \cite{therm42}. The description of macrocanonical ensemble theory can be found in Refs. \cite{ENST01,ENST05,ENST06}. The microcanonical approach was described in Ref. \cite{ENST07}. The main characteristics of these canonical ensemble theories were compared in \cite{therm42}, as shown in Figure \ref{canonical}. In framework of the canonical ensemble theory but within the grand-canonical ensemble limits, the results are similar to the macrocanonical ensemble theory \cite{IBD146}. These different canonical ensemble theories differ mostly in the constraints on mass and charge, and also in the partitions probabilities.

The free energy $F_{f}$ of a partition $f$ is connected to its entropy $S_{f}$ and energy $E_{f}$ via the conventional thermodynamical formulae,
\be
S_f=-(\frac{\partial F_{f}}{\partial T})_{V, \{N_{AZ}\}},~~~~~
E_{f}=F_{f}+TS_{f}.
\ee
For a composite nuclear fragment in a hot nuclear medium, it is very complicate to calculate its free energy $F_{AZ}$. The frequently adopted method is to assume the fragments to be liquid drops which have a spherical shape at the normal nuclear density. The degrees of freedom including the change of fragment shape, density, rotational and vibrational degrees can be easily included. For an individual fragments ($A, Z$) of $A >$ 4, the free energy is parameterized as a form including the bulk ($F^{V}_{AZ}$), surface ($F^{S}_{AZ}$), symmetry ($F^{sym}_{AZ}$) and Coulomb ($E^{C}_{AZ}$) contributions \cite{Ogul17NST},
\be
F_{AZ}=F^{V}_{AZ} + F^{sym}_{AZ} + F^{S}_{AZ} + E^{C}_{AZ}.
\ee
In some cases, the pairing contribution will also be considered \cite{IBD146}. A temperature-dependent parameterized free energy formula will also provide help, such as in the works of \cite{FrEnerg01,FrEnerg02,FrEnerg03}. In a modified Fisher model, the free energy of fragment is parameterized as the Weisz\"{a}cker-Beth semiclassical mass formula with a density and temperature dependent coefficients \cite{IBD144,WBF1,WBF2,IBDLiu1,IBDLiu2,IBDLiu3,IBD147,IBD148,IBD149}.
One can refer to the systematic comparison of thermodynamic models in Refs. \cite{therm42,ENST01}.

\subsubsection{Abrasion-Ablation Models \label{sec:MSAA}}

The abrasion-ablation (AA) model describes the process of heavy-ion collisions in a two-stages manner \cite{AA}. In the first stage, i.e., the colliding stage, the abrasion of nucleons happens in the overlapping zone between projectile and target nuclei, which induces nucleons lost of original nucleus and makes the residues be highly excited due to excitation. The residue nucleus is called as the primary fragment, which has many holes due to the abraded nucleons. In the second stage, i.e., the de-excitation stage, the highly excited residues emit light particles, such as protons, neutrons, or $\alpha$, to be bounded isotopes. The end of this stage corresponds to the chemical freeze-out of the system. The bounded isotope can still emit photons to become to the ground states, but is not considered in the abrasion-ablation models. In the relativistic energy range, the projectile and target nuclei are assumed to be hard balls, the probabilities of nucleon collisions are determined by nucleon-nucleon reaction cross sections. For the intermediate energy collisions, the Pauli effects should be considered. If the projectile or target nucleus is an very asymmetric one, the neutron and proton densities should be much different. The modified statistical abrasion-ablation (SAA) model introduces different neutron and proton density distributions, which improves the predictions for the isotopic distributions. We would like to introduce the modified SAA model.

\begin{figure}[htbp]
\begin{center}
\begin{minipage}[t]{15 cm}
\epsfig{file=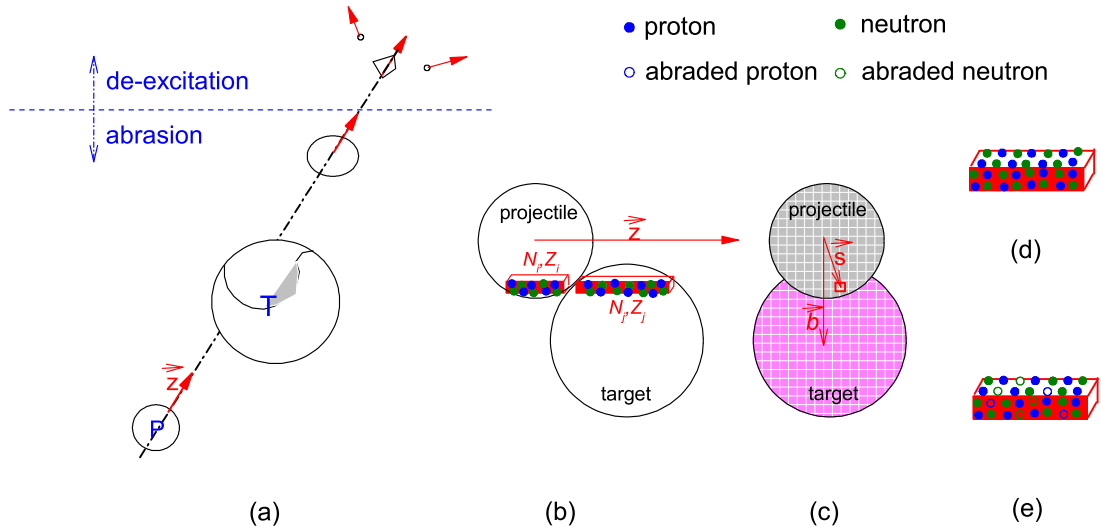,scale=0.8}
\end{minipage}
\begin{minipage}[t]{16.5 cm}
\caption{(Color online) (a) Schematic picture of colliding between the projectile and target nuclei in the statistical abrasion ablation model. (b) Typical overlapping finite tubes for projectile and target nuclei along the beam direction. (c) The overlapping tubes for projectile and target nuclei in the perpendicular direction to beam. (d) The protons and neutrons in the finite tube before abrasion. (e) The protons and neutrons in the finite tube after the abrasion. $\vec{z}$ and $\vec{b}$ denote the beam direction and impact parameter, and $\vec{s}$ denotes the vector from the centre of the projectile nucleus to the tube. \label{saadraw}}
\end{minipage}
\end{center}
\end{figure}

The modified SAA model adopts the same two-stages method as AA model while different functions are used to describe the nuclear densities for protons and neutrons (A more detailed description can be found in Refs. \cite{Ma09PRC,Fang00PRC}). We would like to describe the methods in SAA according to Figure \ref{saadraw}. In Fig. \ref{saadraw} (a), the collisions between the projectile to the target nucleus are plotted, for whom the ``participants'' and ``spectators'' are shown. The overlapping zone between the projectile and target nuclei is abraded, which forms the highly excited primary fragments; and then the primary fragments undergo the second stage of the reaction, which de-excite to form final (cold) fragments by emitting light particles. In the first abrasion stage, the projectile and target nuclei are assumed to be composed of infinitesimal tubes, which are arranged along the beam direction ($\vec{z}$). A typical pair of infinitesimal tubes for the projectile and target nuclei, with neutron and proton numbers ($N_i, Z_i$) and ($N_j, Z_j$), respectively, can be found in the profile along $\vec{z}$ in Fig. \ref{saadraw} (b). In (c), the profiles are shown as grids for the colliding projectile and target nuclei which is vertical to the beam direction is shown, with the infinitesimal tubes for them. $\vec{s}$ and $\vec{b}$ are defined in the plane perpendicular to the beam direction. The numbers of neutrons and protons in the tubes are determined by their density distributions ($\rho_n, \rho_p$). For a symmetric nucleus, neutrons and protons are almost in uniform distribution. While for an asymmetric nucleus, in some tubes, in particular in the skirt zone of nucleus, the neutrons and protons numbers may have large difference. The density distributions of protons and neutrons are described by two different functions in the SAA model. In (d), the protons and neutrons in a infinitesimal tube is shown, and in (e), the protons and neutrons in the tube after the random abrasion in the first stage are shown. The probability of the abrasion is decided by the nucleon-nucleon cross sections.  As results of nucleon abrasions, the information entropy in the tubes in (e) is changed compared to that in (d). The tubes themselves contain the chaoticity of nucleus, which is changed in the abrasion stage. In the following, the formulism for the SAA model to describe the abrasion stage are briefly introduced.

At a given impact parameter $\vec{b}$, for an infinitesimal tube in the projectile, the transmission probabilities for neutrons (protons) are calculated by,
\be
t_{k}(\vec{s}-\vec{b})=\mbox{exp}\{-[D_{n}^{T}(\vec{s}-\vec{b})\sigma_{nk}+D_{n}^{T}(\vec{s}-\vec{b})\sigma_{pk}]\}, \label{eq:saatrans}
\ee
where $D_{n}^{T}$ and $D_{n}^{T}$ are the nuclear density distributions for neutrons and protons of the target nucleus integrated along the beam direction, which are normalized by $\int d^{2}s D^{T}_{n} = N^{T}$ and $\int d^{2}s D^{T}_{p} = Z^{P}$. $N^{T}$ and $Z^{P}$ denotes the neutron numbers $N$ and proton numbers $Z$ of the target nucleus, respectively. $\sigma_{k'k}$ is the nucleon-nucleon cross sections ($k', k = p$ for proton and $k', k = n$ for neutron).

The average absorbed mass in the limit to infinitesimal tubes at a given $\vec{b}$ is,
\be
<\Delta A(b)> = \int d^2sD{_n^T}(s)[1-t_n(\vec{s}-\vec{b})]
+\int d^2sD{_p^T}(s)[1-t_p(\vec{s}-\vec{b})]. \label{eq:Saa-absorb}
\ee
For the projectile spectator, the excitation energy is estimated by $E^* = 13.3 <\Delta A(b)>$, where 13.3 denotes the mean excitation energy for the abraded nucleon from the initial projectile nucleus \cite{AA}. The production cross section for a specific isotope (primary fragment) is calculated by integrating the whole impact parameter range,
\be
\sigma(\Delta N, \Delta Z) = \int d^{2}b P(\Delta N, b)P(\Delta Z, b),\label{eq:SAA-prim}
\ee
with $P(\Delta N, b)$ and $P(\Delta Z, b)$ being the probability distributions for the abraded neutrons and protons at $b$, respectively. For the primary fragment, the total excitation energy are calculated from the abraded protons and neutrons, which will be used to check whether a light particle like proton, neutron, $\alpha$ can be emitted in the second stage of de-excitation.

In the second stage of reaction, the primary fragments will be cooled down by emitting light particles. The excitation energy of a primary fragment will be compared to its separation energy for neutron ($S_n$) or two neutrons ($S_{2n}$), proton ($S_p$) or two protons ($S_{2p}$), and $\alpha$ ($S_{\alpha}$) of the corresponding isotope to decide whether one or two neutrons, one or two protons, or $\alpha$ could be emitted. At the time the excitation energy of a fragment is below the limit to emit more particle, i.e. $E^{*} <$ min($S_n, S_p, S_{\alpha}$), the fragment becomes to the final fragment which is bounded. The final fragment again can be viewed as a droplet though it could emit $\gamma$ particles to decay to the ground state (but not treated in the SAA model).

We would like to address two questions in the SAA model. One is related to $\sigma_{k'k}$. The parameterizations for $\sigma_{k'k}$ in free-space is obtained by fitting the $^{12}$C + $^{12}$C reactions by Charagi and Gupta \cite{FreeSigNN}. When the incident energy of projectile in laboratory reference ($E_{lab}$) is within the range of 10 MeV/u $< E_{lab} <$ 1000 MeV/u,
\bea
\sigma_{pp, nn} &=& \sigma_{pp} =\sigma_{nn} = 13.73 - 15.04\beta^{-1} + 8.76 \beta^{-2} + 68.67\beta^{4}, \label{eq:Freesigmapp}\\
\sigma_{np} &=&  -70.67 - 18.18\beta^{-1} + 25.26 \beta^{-2} + 113.85\beta, \label{eq:Freesigmanp}\\
\beta &=&  v/c = \sqrt{1.0-\frac{1.0}{(1.0+E_{lab}/935.1)^{2}}} \nonumber
\eea
$\sigma$ is in mb, and $E_{lab}$ is in MeV/u. Considering the in-medium effect of $\sigma_{k'k}$, Cai \textit{et al} incorporated the theoretical results based on the Bonn NN potential and the Dirac-Brueckner approach for nuclear matter by Li and Machleidt \cite{SigIMLM1,SigIMLM2}, and provided the parameterizations for nucleon-nucleon cross sections which depend on nuclear density \cite{SigIMCai},
\bea
\sigma_{pp, nn}^{*} &=& \sigma_{pp, nn}\frac{1.0+7.772E_{lab}^{0.06}\rho^{1.48}}{1.0+19.01\rho^{1.46}},\label{eq:Cainn}\\
\sigma_{np}^{*} &=& \sigma_{np}\frac{1.0+20.88E_{lab}^{0.04}\rho^{2.02}}{1.0+35.86\rho^{1.90}}, \label{eq:Cainp}
\eea
in which $\rho = $ 0.17 $fm^{-3}$ is the saturation density of nuclear matter. The parameterizations for $\sigma_{k'k}$ by Cai \textit{et al} are frequently used, which are labeled as $\sigma_{NN}$.

\begin{figure}[htbp]
\begin{center}
\begin{minipage}[t]{10 cm}
\epsfig{file=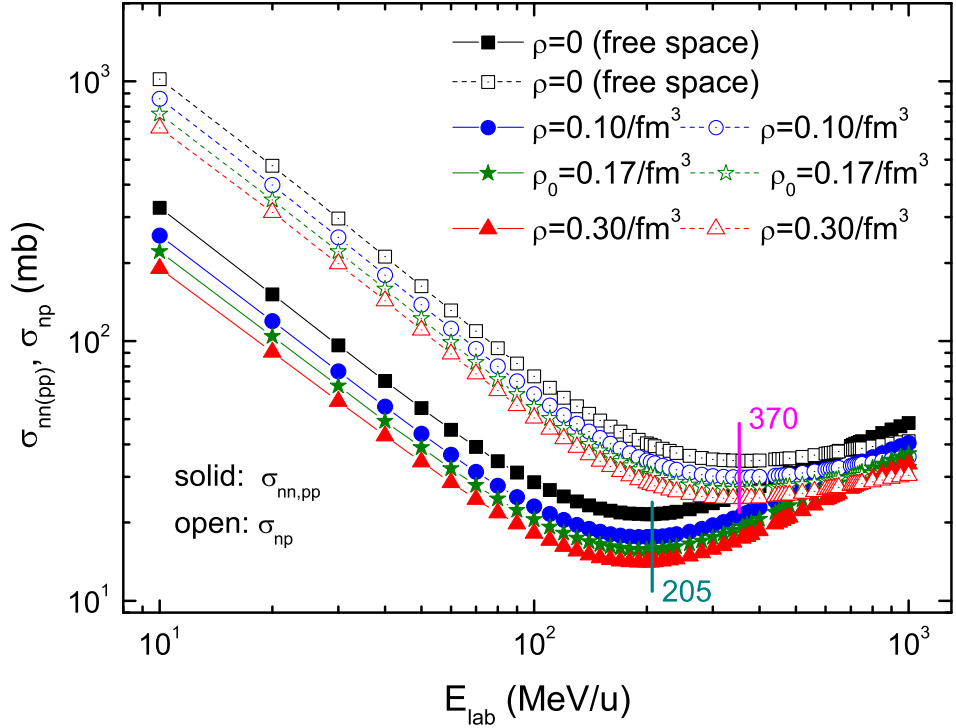,scale=0.8}
\end{minipage}
\begin{minipage}[t]{16.5 cm}
\caption{(Color online) A comparison between the free space and in-medium parameterizations for nucleus-nucleus cross sections. The squares denote the results for the free space ($\rho = $). The circles, stars and triangles denote the results for the in-medium ones at $\rho = 0$ 0.10, $\rho_{0} =$0.17 and $\rho = $ 0.3 fm$^{-3}$, respectively. The full symbols denote the results calculated by Eqs. (\ref{eq:Freesigmapp}) and (\ref{eq:Freesigmanp}) for the free space and the in-medium ones, respectively. The full symbols denote the results calculated by Eqs. (\ref{eq:Cainn}) and (\ref{eq:Cainp}) for the free space and the in-medium ones, respectively. The two bars at 205 MeV/u and 370 MeV/u denote where $\sigma_{pp, nn}$ and $\sigma_{np}$ are the minimums. \label{saaSigFrIM}}
\end{minipage}
\end{center}
\end{figure}

The comparison between the nucleon-nucleon cross sections for the free-space and in-medium at $\rho =$ 0.10, 0.17 ($\rho_0$), and 0.30 fm$^{-3}$, within the range of 10 MeV/u $\leq E_{lab} \leq$ 1000 MeV/u, can be found in Figure \ref{saaSigFrIM}. The minimums for the free-space $\sigma_{pp, nn}$ and $\sigma_{np}$, as well as for the in-medium ones, appear at $E_{lab} =$ 205 and 370 MeV/u, respectively. $\sigma_{pp, nn}$ decrease quickly when $E_{lab} <$ 150 MeV/u. The same phenomena happens for $\sigma_{np}$ when $E_{lab} <$ 200 MeV/u. $\sigma_{pp, nn}$ increases much faster than $\sigma_{np}$, which results in $\sigma_{np} > $ 640 MeV/u.

A comparison between the isotopic yields in the SAA model by adopting the free-space and in-medium $\sigma_{k'k}$ can be found in Ref. \cite{SAAFrIm}, which indicates that for central collisions, it is better to adopt the in-medium $\sigma_{NN}$, while for the peripheral collisions it is better to adopt the free-space $\sigma_{k'k}$. Warner \textit{et al} further analyzed the density dependence of Eqs. (\ref{eq:Cainn}) and (\ref{eq:Cainp}). By considering the Pauli blocking effect of nucleons, they modified the parameterizations of Cai's to \cite{SigWarner},
\be
\sigma_{NN}^{*} = \sigma_{NN}P(\frac{E_{F}^{T}}{E}), ~~~P(x) = 1 - \frac{7x}{5}, \label{eq:SigWarner}
\ee
where $E_{F}^{T}$ is the Fermi energy of target nucleus. The Fermi energy and the nuclear density are correlated via the Fermi momentum and the local nuclear density of target, $k_{F}^{T}(r) = [3\pi^2\rho_{T}(r)/2]^{1/3}$.
A simple comparison for the influence of $\sigma_{k'k}$ on reaction cross section ($\sigma_{R}$) can be found in Ref. \cite{FrIMSig}.

The second question is about the proton and neutron density distributions for asymmetric nucleus, for which the Fermi-type density distribution is usually adopted  \cite{SAA-rho-Fermi},
\be
\rho_{i}(r) = \frac{\rho_{i}^{0}}{1+\mbox{exp}(\frac{r-C_{i}}{f_it_{i}/4.4})}, ~~~~~i = n, p,  \label{eq:SAA-Fermi}
\ee
with $\rho_{i}^{0}$ being a normalization constant which ensures the integration of the density distribution equals the numbers of neutrons or protons. $t_{i}$ and $C_i$ are the diffuseness parameters and the half the density radius of neutron or proton distribution, respectively. For an asymmetric nucleus, $f_i$ can be adjusted to change the diffuseness if there is a neutron skin or proton skin structure \cite{SAA-Fermi-fi,IBD152}.

Comparing the initial projectile nucleus, in the description of cartoon way, the primary fragments and the final fragments, the nucleus change from a liquid drop to many hot drops with holes in it due to abrasion, and then form the many cold liquid drops. Similarly, the information entropy of the system should also experiences the process.

\subsubsection{Macroscopic-Microscopic Models \label{sec:Mhipse}}

The prediction ability of the SAA model is rather limited though it reproduces well the cross section of fragment. For example, it cannot predict the velocity or momentum of the fragments, which are very important to study many process in heavy-ion collisions. The heavy ion phase space exploration ({\sc hipse}) model is proposed to study the reaction processes with few important parameters in the fully microscopic manner based on a macroscopic-microscopic ``phenomenology'', and accounts for both the dynamical and statistical properties of nuclear collisions \cite{Hipse-mocko,Hipse-1}.

The nuclear reaction in the {\sc hipse} model is separated into three stages. The first stage is the approaching of the projectile nucleus to the target nucleus. The second stage is the partitioning of the system, which forms the fragments. The third stage describes cluster propagation, in which an in-flight statistical decay is adopted \cite{Hipse-1}. In the centre of masses reference between the projectile and target nuclei, the classical two-body dynamics is assumed for the entrance channel. At large distance (the projectile is far from the target nucleus), the macroscopic proximity potential is used to describe the nucleus-nucleus potential, and gives a realistic Coulomb barrier. When the distance between the projectile and target nuclei is relative small, the nucleus-nucleus potential becomes sharper with the increasing beam energy. A phenomenological parameter ($\alpha_{a}$), with $\alpha_{a} \leq$ 0 for the adiabatic limit and $\alpha_{a} =$ 1 for sudden approximation, is introduced to extrapolate, respectively. At the minimal distance approach, the sampling of nucleons in each nucleus is according to a realistic zero-temperature Thomas-Fermi distribution, and a simple geometrical considerations is adopted to treat the participant and spectator regions. The quasi-projectile and quasi-target spectators are formed by the nucleons outside the overlap region. Both of direct nucleon-nucleon collisions and nucleon exchange effect should be treated. The direct nucleon-nucleon collision becomes increasingly important as the beam energy increases. The in-medium collisions are considered by modelling the fractions ($\chi_{coll}$) of nucleons in the overlap regions, which slightly distort the Fermi motion hypothesis in the sampling. At the time the direct collisions are over, a fraction of the nucleons, $\chi_{ex}$, in the overlap region is exchanged between the spectator nuclei in the picture of relaxing the pure participant-spectator. The above simulation forms the first stage for {\sc hipse} model. After these preliminary steps, the coalescence algorithm is used to form the clusters according to the classical Hamiltonian with the same nucleus-nucleus potential as in the approach phase. To incorporate the reactions below the Fermi energy, after the freeze-out time ($t_{froz}$), two fragments fuse if the relative separation is less than their fuse barrier distance according to the Coulomb barrier, from which the final state interaction is stopped and the exchange of particle no longer happens. At this stage, the total excitation energy of one cluster can be determined event by event form the law of energy conservation, while in the SAA model the excitation energy is proportional to the abraded nucleons. The clusters at this time can be sent to undergo the in-flight decay process. The phase space generated by {\sc hipse} can be sent to the improved {\sc simon} decay model \cite{simon} or the {\sc gemini} decay code \cite{gemini} to carry out the in-flight decay, for which the {\sc gemini} is believed to more better describe the sequential decays of excited nuclei.
With the least adjustable parameters, $\alpha_{a}, \chi_{ex}$ and $\chi_{coll}$, the {\sc hipse} model has been successfully used to simulate the measured 140 MeV/u $^{40, 48}$Ca/$^{58, 64}$Ni + $^9$Be reactions carried out by Mocko \textit{et al} at National Superconducting Cyclotron Laboratory located at the Michigan State University \cite{Hipse-mocko}.

The time evolution of collision can be studied in the {\sc hipse} model, which avoids the disadvantages that no evolution of system is reflected in the SAA model.

\subsubsection{Abrasion plus Canonical Thermodynamic Model}

Mallik \textit{et al} \cite{AACTM1,AACTM2,FrgDifT01} proposed a model to calculated the cross sections of various reaction products in disintegration of projectile-like fragments, which includes three parts: (1) abrasion, (2) disintegration of hot abraded projectile-like fragment, and (3) possible evaporation of hot primary fragments. The model is organized based on the physics between the phenomenological empirical parametrization of fragmentation cross sections ({\sc EPAX}) and HIPSE model and AMD model. The geometric abrasion model is expanded to include dynamic effects using a transport model \cite{AACTM1}. In the first stage of abrasion, they calculate the abrasion cross section with probability $P_{N_{s},Z_{s}}(b)$ ($N_{s}, Z_{s}$ denote the neutrons and protons in the projectile-like fragment) at an impact parameter $b$. An improvement has been achieved by incorporating an impact-parameter-dependent temperature profile in the abrasion simulation \cite{AACTM2} (the non-uniform temperature for primary fragments and cold fragments have also been shown \cite{FrEnerg03,IBDLiu2,IBDLiu3,FrgDifT01,FrgDifT02,FrgDifT03,FrgDifT04}). In the second stage, the fragment in the first stage undergoes expanding and breaking up process, which turns into many excited composites and nucleons. This stage is dealt with CTM \cite{ENST01}. In the third stage, for each excited composites formed in the second stage, a de-excitation process will be carried out with the help of Monte Carlo simulations \cite{AACTMdecay}. Both the decays of particle emissions and $\gamma$ emission are considered in the de-excitation process. Nice predictions for cross sections of measured fragments have been achieved by this model for the above 140 MeV/u reactions \cite{AACTM1,AACTM2,AACTMdecay,AACTM3,AACTM4}, as well as the isoscaling phenomenon \cite{AACTMdecay} and the bimodality phenomenon \cite{AACTM5}.

\subsection{De-excitation Models}

After the primary fragments are formed, they should undergo a de-excitation process before they are compared to the experimental results. An after-burner or evaporation code, which is always coupled to statistical or dynamical codes, allows the decay of hot fragments to to a bounded nuclei maybe in ground state or bounded states. For the system evolution, the emission of particles in the decay process further alter the chaoticity. It is also important to introduce the generally used models describing de-excitations. In this subsection, we briefly introduce these codes.

\subsubsection{\sc{gemini} Model}

Among these codes, {\sc gemini} \cite{gemini} is the most widely used one, which deals with the decay of compound nuclei in a sequential binary decays manner in the framework of statistical model. A Monte Carlo technique is employed to follow the decay chains of individual compound nuclei through sequential binary decays until the products can no longer undergo further decay. All possible binary divisions from light-particle emission to symmetric division are considered. The Hauser-Feshbach formalism \cite{gemini-HF48} is adopted to calculate the decay width for the evaporation of fragments with $Z \leq$ 2.

The thermal excitation energy of the residue system includes the excitation energy, binding energy, ration and deformation energy, which are related to the spins of the initial system, the residue system, and the emitted particles. In the calculation of binding energy, the Yukawa + exponential model is included \cite{gemini-BE50}, in which the shell and pairing energies are believed to wash out at high excitation energies. The Fermi gas formula is used to deal with the level densities \cite{gemini-Fg53,gemini-Fg54}.

The {\sc gemini++} improved the original {\sc gemini} code with the help of the Bohr-Wheeler fission width \cite{gemini-fis} in conjunction with the systematics of mass distributions \cite{gemini-fismass}, which cures the shortcoming of over prediction of the fission mass and charge distributions of heavy systems.

The binary-decay modes to nucleon and light-nucleus evaporation are allowed in {\sc gemini/gemini++}, i.e., the decaying nucleus can emit a fragment of any mass. This generic binary-decay mode is necessary for describing the complex-fragment formation, and makes the {\sc gemini/gemini++} different to most of the other de-excitation models.

\subsubsection{Statistical Multifragmentation Model \label{sec:SMM}}

Various version of statistical multifragmentation model (SMM) exists \cite{SMMDeComp}, while they are not equivalent. The $N-$body source correlations are assumed to be exhausted by clusterization. These codes differ in the freeze-out volume prescription, the treatment of continuum states and the numerical method for phase space spanning. A systematic comparison for the SMM codes can be found in Ref. \cite{SMMDeComp}.

In general, the SMM models treat the nuclear de-excitation combining the compound-nucleus process at low excitation energies and multifragmentation at high energies \cite{SMM3}. It is assumed that the excited, thermalized nuclear system expands, breaking up simultaneously into several fragments as the volume drops below a certain low-density where meets the freeze-out volume. A thermodynamic weight is assigned to each fragment partition, which is used to choose a multifragmentation partition at random. At low excitation energies, only fragment partitions with total multiplicity smaller than four are considered, which includes the binary and ternary decay channels. At high excitation energies, all available channels are taken into account. The competition with the compound-nucleus channel are considered in SMM, which falls back naturally to conventional evaporation and fission processes at low excitation energy.

The multifragmentation products will undergo de-excitation by conventional method. The $A\leq$ 16 fragments undergo Fermi break-up, while de-excitation of heavy fragments is mainly through particle evaporation and fission. The Weisskopf-Ewing formula \cite{smm9} is adopted to deal with the evaporation process in which the emitted nucleus is up to $^{18}$O, and the Bohr-Wheeler formula \cite{gemini-fis} is adopted to deal with the fission width.

\subsubsection{{\sc ABLA07} \label{sec:abla}}

{\sc ABLA07} is improved to be a dynamical code \cite{abla07} compared to the old version {\sc ABLA} of a pure statistical code. The {\sc ABLA07} code treats the de-excitation of the thermalized system by simulating the break-up, particle emission, and fission processes. In berak-up, the thermal instable hot nucleus is cracked into several fragments. The particle evaporation is dealt by the Weisskopf-Ewing formalism \cite{smm9}, and the dynamical effects is considered in calculating the fission decay width \cite{abla21}.

The light particles, including neutrons, light charged particles ($Z \leq$ 2), intermediate-mass fragments with $Z >$ 2 and $\gamma$ ray are considered in {\sc ABLA07}. The inverse cross sections are calculated based on nuclear potential, which is energy dependent, and are also used to calculate the kinetic-energy spectra of the emitted particles. The Bass potential is used to calculate the barriers for charged particles \cite{abla23}. The thermal expansion of the source is considered \cite{abla24}. The change of angular momentum induced by the particle emission is also considered. And in the decay of hot excited system with an increased volume, a stage of simultaneous break-up is treated \cite{abla09} (but the compression effect in the colliding stage is not considered).

The de-excitation models are frequently used in the simulation of nuclear reactions and the results are usually compared in a wide range of incident energies \cite{SMMDeComp,DeComp1,DeComp2,DeComp3,DeComp4,DeComp5}.

\section{Shannon Information Entropy and Heavy-ion Collisions \label{sec:SHEapp}}

The first use of information entropy in nuclear/particle physics is to study the hadronic productions by Cao and Hwa \cite{BranchRef2}, adopting the general definition of Shannon information entropy. Y.G. Ma adopted an idea of ``event entropy'' to study the liquid-gas phase transition in heavy-ion collisions, which is for the discrete particles \cite{YG99PRL}. G.L. Ma \textit{et al} extended the information entropy for continuum in the analysis and investigated the $\Delta-$scaling phenomenon and the liquid-gas phase transition in a series of reactions \cite{LGT132}. Recently, C.W. Ma \textit{et al} adopted the event information entropy for intermediate mass fragments and found the isobaric scaling phenomenon in neutron-rich projectile fragmentation reactions \cite{Ma15PLB} as well as in the fragments differing of different neutron-excesses \cite{IBD154}. In this section, we review the applications of Shannon information entropy analysis in heavy-ion collisions and the fragment productions in them.

\subsection{Branching Process}

Entropy is an important characteristic of multiparticle production process. The particle production processes are considered as dynamical systems in which the entropy increases generally. Particle production in a branching process occurs by degradation of virtual mass and splitting of a particle into other particles. In high energy collisions, where the question of chaos in gauge dynamics arises, only the particles in final states can be measured. To verify if chaotic behavior happens, it is necessary to quantify the loss of information at the end of the branching process, as well as be presented in a suitable form for experimental determination \cite{BranchRef2}. In quantum chromodynamics (QCD), the particle which can initiate the process may be a quark of a gluon, and in QED the particle may be an electron. In QCD, the gluon distribution is very different from the photon distribution of QED since the non-Abelian nature and the self-interacting gluons \cite{BranchRef1}. The number of degrees of freedom increases with time evolution in  dynamical system. The measurable entropy index was introduced to describe the degree of fluctuation of the final particles event by event \cite{BranchRef2,BranchRef1}.

The following quantity, $V_{i}$, is regarded as an adequate parameter to measure chaotic behavior of branching process in the QCD model with appropriate splitting functions,
\be
V_{i} \equiv (<n^{2}>_{i}-<n>_{i}^{2})/<n>_{i}^{2}, \label{eq:Branch-Vi}
\ee
where $i$ is the number of steps in the branching process and $n$ is the number of produced particles in an event. For large $i$, a relatively $i$ independent $V_{i}$ indicating chaos, for which the decreasing $V_{i}$ with $i$ means the absence of chaos. In QCD chaos occurs while it does not in Abelian theory \cite{BranchQCDVi}. With a probability distribution $P_{i}(n)$ of particles $n$ produced at level $i$, the information entropy $S_{i}$ for the branching process is defined as
\be
S_{i} \equiv -\sum_{n}P_{i}(n)\ln P_{i}(n), \label{eq:Branch-Si}
\ee
This definition of $S_{i}$ was applied to particle production \cite{BranchQCD} and was suggested it as a natural and general indicator for chaoticity in the time evolution of a branching process \cite{BranchRef1}.

\begin{figure}[htbp]
\begin{center}
\begin{minipage}[t]{16.5 cm}
\epsfig{file=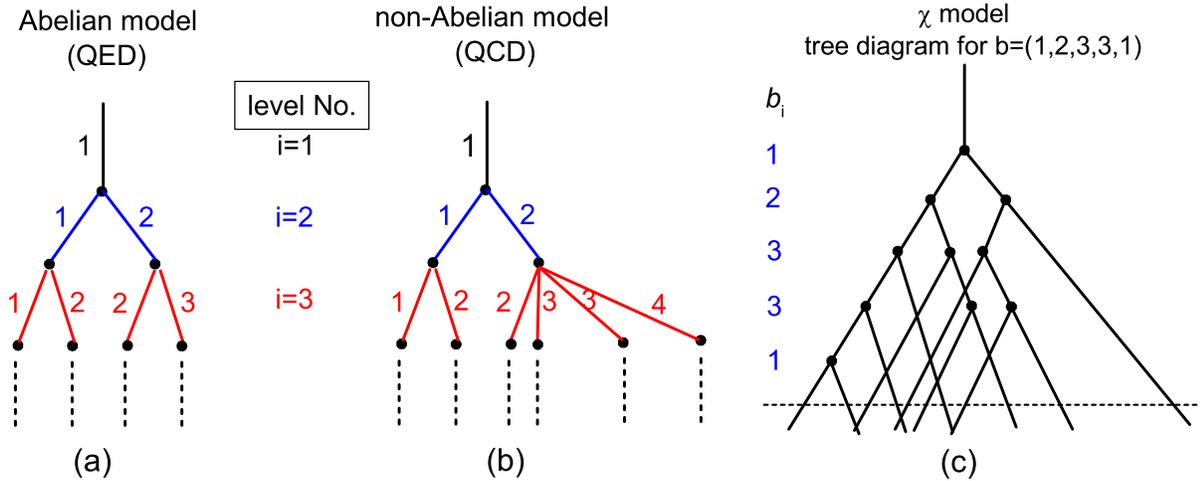,scale=1.1}
\end{minipage}
\begin{minipage}[t]{16.5 cm}
\caption{(Color online) Tree diagrams for branching process in Abelian model [in (a)] and non-Abelian model [in (b)] in Ref. \cite{BranchRef1}. $n$ means ($n$ - 1) photons and 1 electron in QED, and means the number of gluons in QCD. In panel (c) a tree diagram for $\vec{b} =$ (1, 2, 3, 3, 1) denotes a branching process in $\chi$ model \cite{BranchRef2}.
\label{Branch}}
\end{minipage}
\end{center}
\end{figure}

Limited to the time evolution in the branching process and focused on the event topology, Brogueira and Deus started with the process of $P_{1}(1) =$ 1 and $S_{1} =$ 0 for QED and pure QCD. The tree diagram for the processes are plotted in Figure \ref{Branch} (a) and (b) for QED and QCD, respectively. In QED, the probability for an electron to remain as an electron is (1$-\alpha$), and it be $\alpha$ to become an electron and a noninteracting photon. The generated probabilistic branching tree can be seen in Figure \ref{Branch} (a). In QCD, the probability for a gluon to remain a gluon is (1$-\alpha$) and the probability for a gluon to become two gluons is $\alpha$. The corresponding generated probabilistic branching tree is shown in Figure \ref{Branch} (b).

In the continuous approximation, the information entropy [Equation (\ref{eq:Branch-Si})] can be obtained,
\be
S_{i}-\ln<n>_{i} = - \int \psi_{i}(z)\ln\psi_{i}(z)dz,
\ee
with $\psi_{i}(z)$ being the Koba-Nielsen-Olesen (KNO) function. In the case of QED, there is no KNO limiting function and $\psi_{i}$ approaching a $\delta$ function \cite{KNO1}. In the case of QCD a limiting KNO function exists:
\be
\psi_{i}(z)_{\overrightarrow{~~~~~i, <n>_{i}\rightarrow \infty~~~~~}}\psi(z),
\ee
which agrees with the original QCD branching calculations \cite{KNO2,KNO3,KNO4}.

\begin{figure}[htbp]
\begin{center}
\begin{minipage}[t]{15.5cm}
\epsfig{file=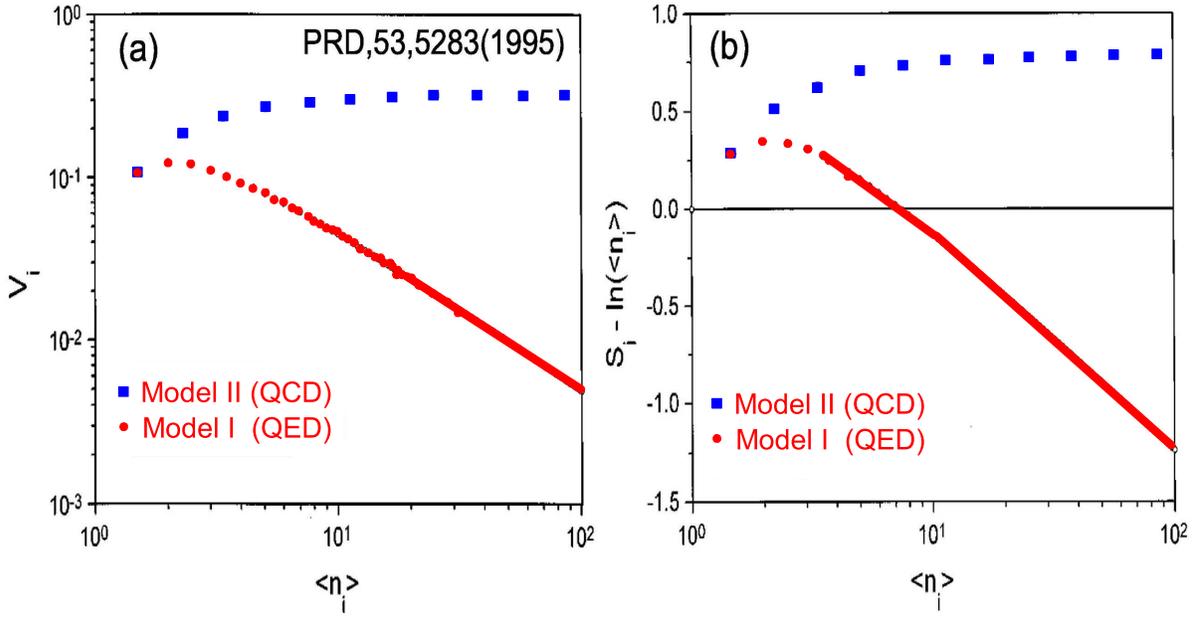,scale=0.85}
\end{minipage}
\begin{minipage}[t]{16.5 cm}
\caption{(Color online) Re-plotted figures for $V_{i}$ and $Q_{i} \equiv S_{i} - \ln<n>_{i}$ from Ref. \cite{BranchRef1}. (a) Parameter $V_{i}$ [Equation (\ref{eq:Branch-Si})] as a function of $<n>_{i}$ for the QED and QCD models, respectively. (b) The quantity $Q_{i}$ [Equation (\ref{eq:qiui})] as a function of $<n>_{i}$ for the QED and QCD models, respectively. \label{QiEntropy}}
\end{minipage}
\end{center}
\end{figure}

The quantity
\be
Q_{i} \equiv S_{i} - \ln<n>_{i} \label{eq:qiui}
\ee
was proposed to measure the evolution of a branching process. The results of $V_{i}$ and $Q_{i}$ at the same situations for the QED and QCD cases can be found in Figure \ref{QiEntropy}(a) and (b), respectively. It can be seen that $Q_{i}$ well reproduce the trends of $V_{i}$ both in the QED and QCD distributions. In the QCD case, $Q_{i}$ approaches a finite constant as $i \rightarrow \infty$ and the process is chaotic. In the QED case, $Q_{i}$ continuously decreases and approaches $-\infty$ as $i \rightarrow \infty$, and the process is not chaotic \cite{BranchRef1}. This conclusion is also supported by the work of Cao and Hwa \cite{BranchRef2,BranchQCDVi}.

The details of the work showing signs indicative of chaos in the perturbative QCD branching, whereas the model lacking the characteristics of QCD shows no signs of chaos \cite{BranchQCDVi}, have been discussed in a following work \cite{BranchRef2}. We briefly introduce the simulation process in Ref. \cite{BranchRef2}. The QCD dynamics by only considering the pure gauge theory without quarks, and a cascade model called as the $\chi$ model which has none of the QCD features have been adopted to generate events of particle production through branching. The same branching processes for the initial parton are considered, i.e., the initial parton has virtuality $Q^2$ and the successive branchings continue until the virtualities of all partons are smaller than $Q_{0}^{2}$. The splitting function at each vertex of branching is
\be
P(z) = 6 \left[\frac{1-z}{z}+\frac{z}{1-z}+z(1-z)\right], \label{eq:splittingQCD}
\ee
in which $z$ is the momentum fraction of the daughter parton in the frame where the father parton's momentum is 1. Only the last term of Equation (\ref{eq:splittingQCD}) is kept in the $\chi$ model, i.e.,
\be
P(z) = 6z(1-z) \label{eq:splittingChi}
\ee
No divergences happen at $z =$ 0 and 1 in the $\chi$ model. While the divergences are the source of complication for QCD, which need to be treated carefully. The dynamics of branching in the $\chi$ model is very different from QCD, and was used to exemplify the Alelian dynamics which has no infrared and collinear divergences (instead of QED in Ref. \cite{BranchRef1}).

Considering no recombination of partons, a tree diagram denotes one branching process. The vertices of the tree could be ordered vertically in accordance to $q^2$ of the father partons. There should be many diagrams with the same topology but describe different evolution processes. The results lead to the same number of particles at the end but belong to different final states. Only the topology of the tree diagram shall be considered if the momenta of the final particles can be deferred in consideration. In this manner the fluctuation of the particle multiplicity can be studied without their momenta, and the trajectory can be defined in the multiplicity space for a branching process.

\begin{figure}[htbp]
\begin{center}
\begin{minipage}[t]{11.5cm}
\epsfig{file=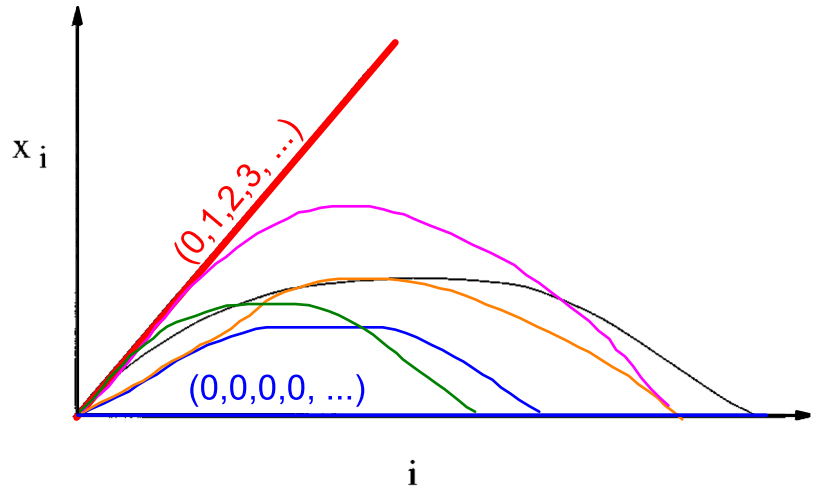,scale=1.0}
\end{minipage}
\begin{minipage}[t]{16.5 cm}
\caption{(Color online) Some possible trajectories $\vec{x}$. \label{Xichitraj}}
\end{minipage}
\end{center}
\end{figure}

An example topology of diagram is plotted in Figure \ref{Branch} (c), in which all partons of the same generation is placed at the same level regardless of their $q^2$. The branching points of the same generation are placed at the same level in the vertical direction. While the partons in the horizontal direction have no significance in the diagram. All partons reaching the final state have $q^2 \leq Q_{0}^{2}$. The generation, denoted by $i$, starts with $i =$ 0 for the initial parton with $q^2 \leq Q^{2}$.  $b_{i}$, which denotes the number of branching points at the $i^{th}$ generation, is defined as a vector $\vec{b} = (b_{0}, b_{1}, b_{2}), ...$ with many components. If $\vec{b}$ is applied to describe an electron radiating photons, a bremsstrahlung diagram would be $\vec{b} =$ (1, 1, 1, ...). While for a cell subdivides into 2, i.e., the cell reproduction, $\vec{b} =$ (1, 2, 4, 8, ...). The example in Figure \ref{Branch} (c) has $\vec{b} = (1, 2, 3, 3, 1)$. This description can be further simplified as
\be
x_{i} = \log_{2}b_{i}, \label{XiDef}
\ee
and the corresponding vector $\vec{x} = (x_{0}, x_{1}, x_{2}, ...)$. In Figure \ref{Xichitraj}, some possible trajectories $\vec{x}$ are plotted. The minimum and maximum of $x_{i}$ are 0 (blue line on Figure \ref{Xichitraj}) and 1 (red line on Figure \ref{Xichitraj}), respectively. The extreme vectors for $x_{i}$ are $\vec{x} =$ (0, 0, 0, 0, ...) and (0, 1, 2, 3, ...), which correspond to the bremsstrahlung and cell reproduction, respectively. All possible tree diagrams of branching processes are represented by a line between the boundary lines (the blue and red lines), which are denoted by thin lines in Figure \ref{Xichitraj} and specify trajectories.

In the final state the complete information about the parton identified as observable particles is registered by their momenta. The quantum fluctuation causes the fluctuation in the final-state momenta, and entropy shall be non-zero, which means that some information is lost. Considering that $p_{j}$ denote the fraction of particles in an event falling into the $j^{th}$ bin of size $\delta$, in the limit of many such bins in the system under the entropy is defined by
\be
S = -\sum_{j}p_{j}\ln p_{j},  \label{eq:SysEntrpy}
\ee
with the information dimension $D_1$ is
\be
D_{1} = -\lim_{\delta\rightarrow\infty}S/\ln\delta
\ee
It should be noted that insertion of zeros between two points on the multiplicity scale of different bins does not influence the entropy, while the shape of $p_{i}$ is changed.

In the following we go to the more detail of the deduction. One who is not interested in this can jump to the conclusions directly. If one uses $z_{i}$ to denote the momentum fraction of an $i^{th}$ generation daughter parton at a branching vertex (the momentum of father parton is 1), the momentum fraction of a final particle is
\be
x = \prod_{i}z_{i}.
\ee
The product is taken over all generations of a particular path in the branching tree, which ranges from the initial parton to the final particle under consideration. A cumulative variable $X$ can also be defined as,
\be
X(x) = \int_{x1}^{x}\rho(x')dx'/\int_{x1}^{x2}\rho(x')/dx',
\ee
with $\rho(x)$ being the averaged results over many events. In the $X$ space the interval 0 $ \leq X \leq$ 1 can be divided into $M$ bins of width $\delta = 1/M$. The factorial moment of $q^{th}$ order is
\be
f_{q}(M) = M^{-1} \sum_{j=1}^{M}n_{j}(n_{j}-1)\cdot\cdot\cdot(n_{j}-q+1),
\ee
in which $n_{j}$ is the number of final particles in the $j^{th}$ bin in any given event. The normalized factorial moment after averaging over all events is
\be
F_{q} = <f_q>/<f_1>^q
\ee

In the event by event analysis, large fluctuations for the produced particles exist in the $X$ space. It is useful to register the fluctuation from event to event in the $X$ distribution, and $F_{q}^{e}$ can be viewed as the horizontal moments in the $X$ space. The vertical moments in the event space can be defined as follows:
\be
<F_{q}^{p}>=\frac{1}{\mathcal{N}}\sum_{e=1}^{\mathcal{N}}(F_{q}^{p})^{p},
\ee
where $\mathcal{N}$ is the total number of events and $p$ is a positive real number. The normalized moments are
\be
C_{p,q}(M)=<F_{q}^{p}(M)>/<F_q(M)>^p,
\ee
If $C_{p,q}(M)$ depends on $M$ in the form of power-law, i.e.,
\be
C_{p,q}(M)\propto M^{\varphi_{q}(p)},
\ee
the entropy index was defined as \cite{BranchRef2},
\be
\mu_{q}=\frac{d}{dp}\varphi_{q}(p)|_{p = 1}.
\ee
Let $P_{q}^{e}$ denote the normalized factorial moment of the $e^{th}$ event:
\be
P_{q}^{e} = F_{q}^{e}/\sum_{e=1}^{\mathcal{N}}F_{q}^{e}.
\ee
In the event space, a new entropy can be defined as,
\be
S_{q}=-\sum_{e=1}^{\mathcal{N}}P_{q}^{e}\ln P_{q}^{e},
\ee
and define the moments,
\be
H_{p,q}=\sum_{e=1}^{\mathcal{N}}(P_{q}^{e})^{p}.
\ee
$H_{p,q}$ is related to $S_q$ by
\be
S_q = -\frac{d}{dp}\ln H_{p,q}|_{p=1}  \label{eq:eventropy}
\ee
$H_{p,q}$ is related to $C_{p,q}$ by
\be
C_{p,q}=\mathcal{N}^{p-1}H_{p,q}.
\ee
One then has,
\be
\frac{d}{dp}\ln C_{p,q}|_{p=1} = \ln\mathcal{N}-S_q
\ee
and
\be
S_q=\ln(\mathcal{N}M^{-\mu_q}) \label{eq:SqtoMuq}
\ee
The definition of $S_q$ is clearly different from the entropy $S$ defined in Equation (\ref{eq:SysEntrpy}). $S_q$ is called as ``eventropy'' to emphasize that it is defined in the event space.

\begin{figure}[htbp]
\begin{center}
\begin{minipage}[t]{13.5cm}
\epsfig{file=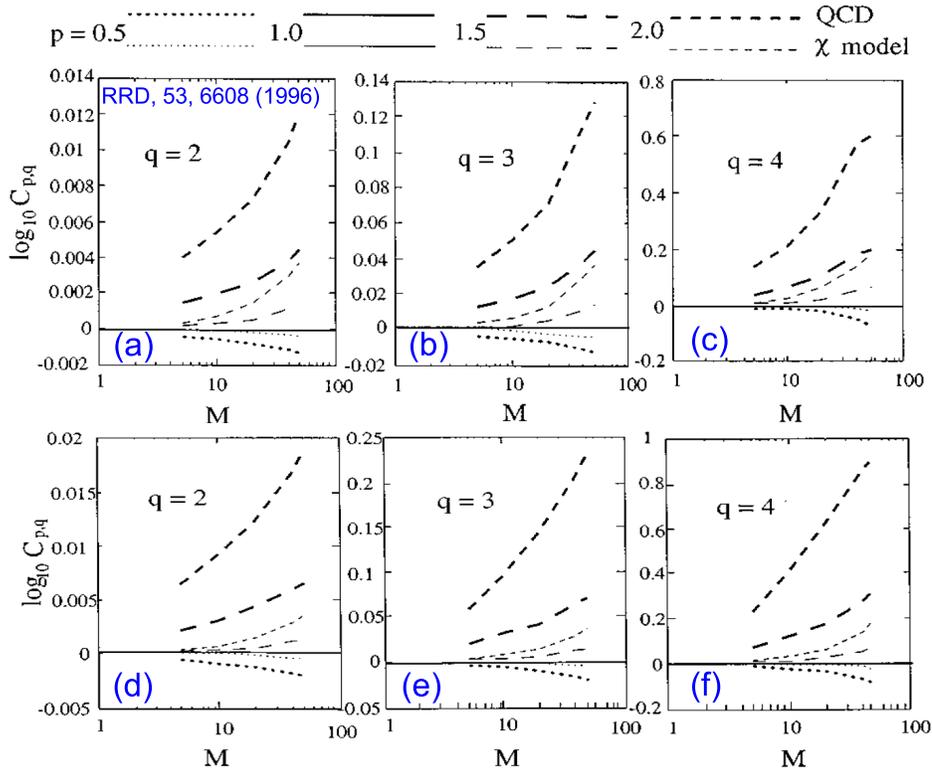,scale=0.95}
\end{minipage}
\begin{minipage}[t]{16.5 cm}
\caption{(Color online) Top: Moments of moments $C_{p, q}$ vs $M$ for $p =$ 0.5-2.0 and $q =$ 2 (a), 3 (b), 4 (c), respectively. Bottom: The same as the top panels but for all events simulated. \label{ChaoticChi}}
\end{minipage}
\end{center}
\end{figure}

At present one can discuss the entropy both for the system and the eventropy, and reaches some conclusions. In the event space, which can be viewed as a one-dimensional space with $\mathcal{N}$, at each site a number $F_{q}^{e}$ can be registered. If $F_{q}^{e}$ is the same at each site, one has $P_{q}^{e} = 1/\mathcal{N}$ and $S_{q} = \ln\mathcal{N}$. In event space, it represents highly disorder since $F_{q}^{e}$ is spread out uniformly over all space. The eventropy will become larger as the number of events increases, which is similar to the situation for $S = \ln M$ where the bin multiplicity is uniformly distributed in the $X$ space of $M$ bins. A branching dynamic that results in the same $F_{q}^{e}$ for every event has no fluctuate in the branching processes, which corresponds to nearby trajectories staying nearby throughout. It means that the dynamics is not chaotic. For $S_q$ to be $\ln \mathcal{N}$, one can see from Equation (\ref{eq:SqtoMuq}) that $\mu_q$ vanishes, which indicates that small $\mu_q$ corresponds to large eventropy and in turn implies no chaotic behavior. In contrary, considering the other extreme, i.e., all $F_{q}^{e} =$ 0 except one event $e'$, then $P_{q}^{e} = \delta_{ee'}$ and $S_q =$ 0. In the event space this is in high order, whereas the fluctuation of $F_{q}^{e}$ from zero to nonzero value is large. In a general speaking, if the distribution of $P(F_{q})$ is a broad one, the fluctuation is large, initially nearby trajectories become widely separated in the final states of different events, and the dynamic is chaotic. To have a small eventropy, $\mu_q$ must be large. One then reaches the conclusion that large entropy index implies chaotic behavior.

In Figure \ref{ChaoticChi}, the correlation between $C_{p, q}$ and $M$ for $p =$ 0.5 - 2.0 and $q =$ 2 , 3, 4 are shown in panels (a), (b), and (c) in double log plots, respectively. $\mu_q$ was determined within the region of $M =$ 5 - 20, for which $C_{p, q}$ for an incremental region of $p$ around 1.
\bea
\mu_q^{(QCD)} &=& 0.0027,~~~0.026,~~~~ 0.15 ~~~~~ (q = 2, 3, 4),  \\
\mu_q^{(\chi)} &=& 0.0011, ~~~0.0087,~~~0.038 ~~~ (q = 2, 3, 4).
\eea
$\mu_q^{(QCD)}$ is significantly larger than $\mu_q^{(\chi)}$, which suggests that the QCD dynamics is a chaotic one while the $\chi$ model is not. For all the events simulated, the new values of $\mu_q$ become to
\bea
\mu_q^{(QCD)} &=& 0.0043,~~~0.044, ~~~~0.20 ~~~~ (q = 2, 3, 4),  \\
\mu_q^{(\chi)} &=& 0.0010,~~~ 0.0089,~~~ 0.039 ~~~ (q = 2, 3, 4).
\eea
The values of $\mu_q^{(QCD)}$ change a lot, but $\mu_q^{(\chi)}$ are nearly the same as the ones within the limited $M$. The difference between $\mu_q^{(QCD)}$ and $\mu_q^{(\chi)}$ becomes even more larger when all events are included, which further supports the conclusion the QCD dynamics is chaotic. Moreover, the entropy indices provide also an efficient way to identify it is a quark or a gluon for the dominant initiator of perturbative QCD branching process \cite{Hwa96PRD}.

To summarize this analysis, one studies the phase space, as well as the event space to extract the entropy indices $\mu_q$. It is also suggested that one can consider a two-dimensional space, for example in the pseudorapidity space as the horizontal axes which have $M$ bins, and the vertical axes has $N$ site corresponding to the $N$ events in the event space. The factorial moments in the pseudorapidity space can be calculated event by event, for which the factorial moments fluctuate greatly. The entropy index $\mu_q$ describes the degree of such fluctuation from event to event. A small and large entropy index $\mu_q$ implies no chaotic and a chaotic behavior, respectively, which suggests that $\mu_q$ is an adequate parameter to measure the chaotic behavior of multiparticle production.

\begin{figure}[htbp]
\begin{center}
\begin{minipage}[t]{14.5cm}
\epsfig{file=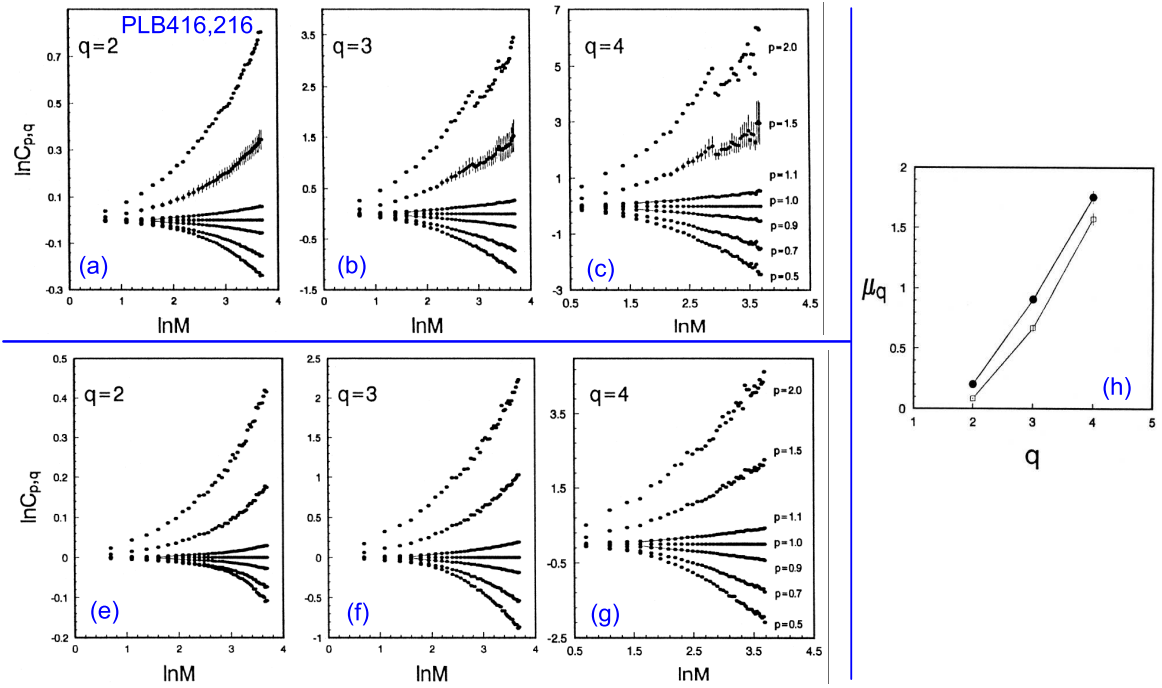,scale=0.95}
\end{minipage}
\begin{minipage}[t]{16.5 cm}
\caption{(Color online) Top: $\ln C_{p, q}$ vs $\ln M$ for $N_{ev} \geq$ 4 event samples at $q =$ 2 (a), 3 (b), 4 (c) measured by NA27 collaboration. Bottom: The same as the top panels but for $N_{ev} \geq$ 10. (h) shows the correlation of $\mu_q$ vs. $q$ determined from the $\ln C_{p, q}$ vs $\ln M$ correlation.
\label{NA27MuQ}}
\end{minipage}
\end{center}
\end{figure}

Wang used the entropy indices to study the $p$ + $p$ collisions at 400 GeV/c measured by the CERN NA27 collaboration \cite{NA27ZPC}. The measured charged particles, with a total number of 3730 non-single-diffractive events ($n_{ch} \geq$ 4) and total number of 2145 events for $n_{ch} \geq$ 10, have been analyzed to determine the entropy indices $\mu_q$ \cite{Wang98PLB,Wang98PRD}. The correlations for $\ln C_{p,q}$ and $\ln M$ of $n_{ch} \geq$ 10 within the region of pseudorapidity (-2 $\leq \eta \leq$ 2) are plotted in Figure \ref{NA27MuQ} (a) ($q =$ 2), (b) ($q =$ 3), and (c) ($q =$ 4), respectively. The same results for the $n_{ch} \geq$ 10 events are plotted in Figure \ref{NA27MuQ} (d), (e), and (f), respectively. $\mu_q$ obtained for $q = $ 2, 3, 4 for the $n_{ch} \geq$ 4 (full circles) and $n_{ch} \geq$ 10 (open squares) events are shown in Figure \ref{NA27MuQ} (h). In reactions dominated by QCD parton showering, the entropy indices increase with decreasing average multiplicities of final states, which indicates a chaotic behavior in the $pp$ reactions. The results of the entropy indices show that, at lower energy, the dominant original parton of the QCD branching process may be a quark for $pp$ collisions.

The physical interpretations of $\mu_q$ as an entropy index are summarized as follows by Ghosh {\it et al} \cite{QuanChaos},
\begin{itemize}
  \item  $\mu_q$ is an appropriate parameter to measure the chaotic behavior of particle production.
  \item  $\mu_q$ describes the degree of fluctuation of the scaled factorial moments in event space as well as the spatial pattern of the particles in the final states, and $\mu_q$ characterizes the degree of fluctuation of the parton multiplicity that initiates branching.
  \item  $\mu_q$ provides an efficient way to  distinguish whether the dominant initiator of the perturbative QCD branching process is a quark or a gluon.
  \item  A small $\mu_q$ implies no chaotic behavior, while a large $\mu_q$ implies chaotic behavior.
\end{itemize}

The chaoticity analysis was applied to multiparticle production in hadronic collisions of $\pi^{+} + p$ and $k^{+} + p$ at 250 GeV$c^{-1}$ \cite{pipkp}, $^{32}$S-AgBr at 200$A$ GeV \cite{SAgBr1,SAgBr2,SAgBr3}, $^{28}$Si-AgBr \cite{SiAgBr1,SiAgBr2}, phase transition \cite{PhsTrs}, and heartbeat irregularity \cite{HrtIrr}.
In the same article, Ghosh {\it et al}. \cite{QuanChaos} commented that the notion of chaotic behavior for particle production in branching processes is more in classical viewpoint than in the quantum one since the concept of trajectory was adopted. While in the quantum viewpoint, the final state of the branching process will vary from event to event though all events start out at precisely the same virtuality. They proposed that, instead of studying the distance between two neighbouring trajectories, it is better to consider all events and examine the mean deviation from the average of the parton multiplicities. They stressed to study of the fluctuations of observable quantities, and used $\mu_q$ to characterize the ``spatial'' properties of the chaotic behavior of multiparticle production process in the hadron-nucleus interaction data of $\pi^{-}-$AgBr at 350 GeV/c measured at CERN. Both the gray and shower particles per event in a collisions are combined as a new parameter, which is named as ``compound multiplicity'', and plays an important role in understanding the reaction dynamics in high-energy nuclear interaction. The chaotic behavior of compound hadrons produced in high-energy hadron-nucleus interactions also offers a unique opportunity to learn about the space-time structure of a strong interacting process. They had reached the following conclusions: 1) The values of the entropy indices for different event samples are positive and quite large compared to the values for the randomized data. 2) Erraticity behavior depends strongly on multiplicity. 3) $\mu_q$ increases with decreasing average multiplicity. And 4) $\mu_q$ for shower particles (pions) falls more rapidly with the average multiplicity than that for compound hadrons.

\subsection{Projectile Fragmentation Reactions \label{sec:SIEHIC}}

In heavy-ion collisions, similar to the branching process, the production of light particles and intermediate isotopes become important. The projectile and target nuclei of reaction system, which can be seen as charged liquid drops, experience violent collisions and are crashed into pieces of different sizes. The multiplicity of the charged particles produced in heavy-ion collisions becomes the simplest observable in experiment, which plays an important role in extracting the first information on the particle production mechanism. It is also important to investigate the multiplicity fluctuations and correlations to understand the internal dynamics of multiparticles production process in reactions induced by heavy-ions from the intermediate energy (above 10 MeV/u \footnote{The frequently used units of incident energy are MeV/u and $A$ MeV, which both represent MeV per nucleon. Both of them are used in this review since in different articles they are not uniformly used.}) to relativistic energy. The question of liquid-gas transition (LGT) in heavy-ion collisions is one of the most important questions around the 1990's (even till now to study the phase transition). The multiplicity of intermediate mass fragment (IMF), namely $N_{imf}$, rises with the beam energy, reaches a maximum and finally falls to lower value. This rise and fall of $N_{imf}$ indicate a LGT in nuclear matter \cite{Ogi91PRL,Tsang93PRL,YG95PRC}. But it is not enough to determine an LGT by the distribution of $N_{imf}$ in heavy-ion collisions. Y.G. Ma first time introduced the Shannon information entropy to study the possible indicator for LGT in heavy-ion collisions \cite{YG99PRL}.
Later, C.W. Ma \textit{et al} applied the Shannon information entropy to analyze the isobaric scaling phenomenon in fragments produced in the neutron-rich heavy-ion collisions \cite{Ma15PLB,IBD154,IBD155}.

In this section, we first introduce the work by Y.G. Ma and others to study the LGT. Then the application of Shannon information entropy for isobaric scaling will be discussed. At last, we look forward to the future applications of information entropy in heavy-ion collisions.

\subsubsection{System Information Entropy as Liquid-Gas Transition Indicator \label{sec:LGT}}

The mass distribution of IMFs indicates a power law with a parameter $\tau$. The minimum of $\tau$, labelled as $\tau_{min}$, appears at the point of liquid-gas transition. But it cannot serve as an essential condition of liquid-gas phase transition since $\tau_{min}$ also appears at supercritical densities along the Kert\'{e}se line \cite{LGT92} and at some subcritical densities at lower temperature \cite{LGT93}. This makes it unclear to determine the phase transition from $N_{imf}$ and $\tau_{min}$.

The nuclear caloric curve can be measured in experiments, which reveals the change of nuclear temperature with excitation energy. The Albergo thermometer is one of the thermometers based on the emission of light particles \cite{TLGT94}. The typical Albergo H-He thermometers is written as,
\be
T_{HHe} = \frac{14.3}{\ln [\frac{Y(^{2}H)\cdot Y(^{4}He)}{Y(^{3}H)\cdot Y(^{3}He)}\cdot 1.6]}, \label{eq:AlbgTHHe}
\ee
where $Y$ denotes the yield of the particle. Experimentally, Pochodzalla \textit{et al} extended it to the frequently used He-Li thermometer \cite{LGT95},
\be
T_{HLi} = \frac{16}{\ln [\frac{Y(^{6}Li)\cdot Y(^{7}Li)}{Y(^{3}He)\cdot Y(^{4}He)}\cdot 2.18]}. \label{eq:AlbgTHeLi}
\ee
The constants in Equations (\ref{eq:AlbgTHHe}) and (\ref{eq:AlbgTHeLi}) changes because they are related to the binding energies and spins of the isotopes. In the general form, the double isotopic ratio temperature can be deduced at the stage of chemical and thermal equilibrium state of the system following the Albergo formula \cite{TLGT94}, which can be found in the work by Tsang \textit{et al} \cite{TLGT98}.

\begin{figure}[htbp]
\begin{center}
\begin{minipage}[t]{16cm}
\epsfig{file=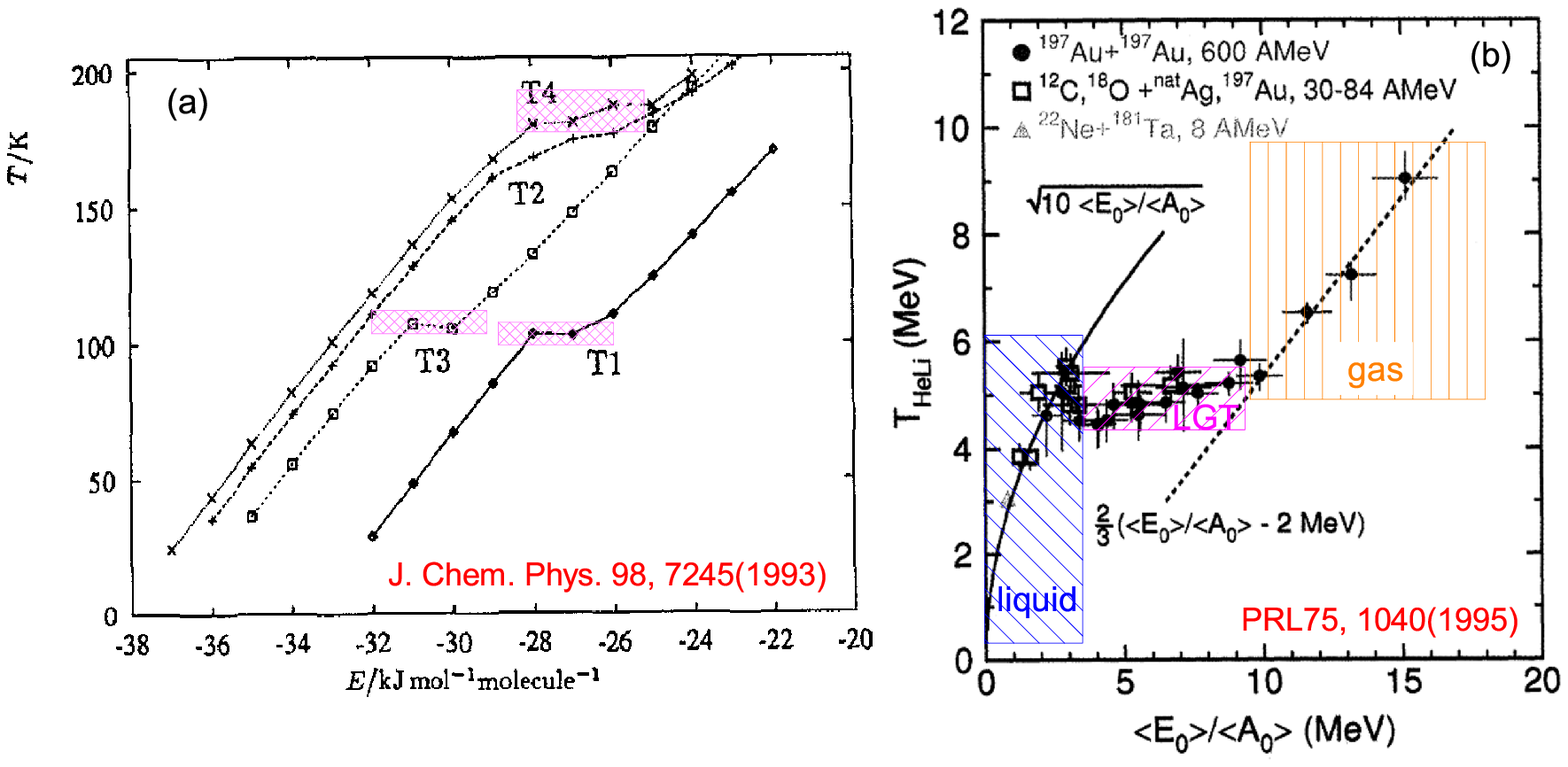,scale=0.9}
\end{minipage}
\begin{minipage}[t]{16.5 cm}
\caption{(Color online) (a) The correlation between temperature and total energy (E) of water simulated for $(H_2O)_8$ \cite{LGT96Water}. (b) The correlation between $T_{HeLi}$ and $\frac{<E_0>}{<A_0>}$ for intermediate energy heavy-ion collisions of 600 MeV/u $^{197}$Au + $^{197}$Au, 30--84 MeV/u $^{12}$C, $^{18}$O + $^{nat}$Ag, $^{197}$Au, and 8 MeV/u $^{22}$Ne + $^{181}$Ta reactions \cite{LGT95}. The three zones marked by different shadows are added to denote the liquid, liquid-gas coexistence, and gas phases, respectively.
\label{LGTTtoExc}}
\end{minipage}
\end{center}
\end{figure}

The $T_{HeLi}$ of IMFs have been measured in the 600 MeV/u $^{197}$Au + $^{197}$Au, 30 -- 84 MeV/u $^{12}$C, $^{18}$O + $^{nat}$Ag, $^{197}$Au, and 8 MeV/u $^{22}$Ne + $^{181}$Ta reactions \cite{LGT95}. It is interesting to find that the caloric curve of nuclei determined by the dependence of $T_{HeLi}$ on the excitation energy per nucleon ($\frac{<E_0>}{<A_0>}$) increases with $\frac{<E_0>}{<A_0>}$ in the ranges of both small ($\frac{<E_0>}{<A_0>} \leq$ 3 MeV) and large ($\frac{<E_0>}{<A_0>} \geq$ 10 MeV), while a plateau of $T_{HeLi}$ is formed within the range of 3 MeV $\leq \frac{<E_0>}{<A_0>} \leq$ 10 MeV (See Figure \ref{LGTTtoExc}).

For water $(H_2O)_8$, in a simulation performed by Wales and Ohmine \cite{LGT96Water}, in the correlation between temperature ($T$) and total energy ($E$), it is clearly shown in Figure \ref{LGTTtoExc} (a) that to melt a $(H_2O)_8$ cluster, a coexistence state of liquid and gas (denoted by the plateau) happens between the liquid and gas phases. Like the same phenomenon illustrated in water, three regions on the caloric curve of $T_{HeLi}$ correspond to the liquid, liquid-gas coexistence, and gas phases, respectively. The plateau was taken as a sign for a first order nuclear liquid gas phase transition. While it was questioned since the change of mass for the Au spectators with excitation energy and side-feeding effect to measure He-Li isotopic temperature \cite{TLGT98,LGT97,LGT99}. It is also criticized that there is a difference between the measured ``apparent'' temperature ($T_{HeLi}$) and the ``real'' temperature, which makes it difficult to perform the direct comparison between them, and hinders people to know the real nuclear caloric curve. Meanwhile, the sharp signature of the liquid-gas phase transition in macroscopic system may be smoothed and blurred due to the small numbers of nucleons in nuclei \cite{YG99PRL}.

Campi \cite{LGT106,LGT109} and Bauer \cite{LGT107} \textit{et al} suggested to use the methods in percolation studies to study the nuclear multifragmentation data. In percolation theory, a signature of critical behavior is contained in the moments of the cluster distribution \cite{LGT108}. The critical exponents were also suggested as an indicator for the liquid-gas transition. The multiplicity, $m$ ($=m_F$ + No. of released protons) of fragments was assumed to be a linear measurement of the distance from the critical points \cite{LGT109}. The region in $m$ below the assumed critical multiplicity $m_c$ is designated as the ``liquid'' phase and that above $m_c$ as the ``gas'' phase. In the liquid phase the heaviest fragment $Z_1$ is omitted in forming the moments, but is not omitted when in the gas phase. Gilkes \textit{et al} stated that the critical exponents for large systems are given in term of the multiplicity difference, $m-m_c=\zeta$, by
\bea
&M_{2}^{*}\sim|\zeta^{-\gamma}|  \label{eq:lgtM2}\\
&Z_1\sim|\zeta|^{\beta}     \label{eq:lgtZ1}\\
&n_z\sim Z^{-\tau} ~~~~\mbox{for}~~~~ m=m_c  \label{eq:lgtnz}\\
\eea
The exponents are related in the form of \cite{LGT110}
\be
\tau=2+\frac{\beta}{\beta+\gamma} \label{eq:lgtExpcorr}
\ee
The second moments are used to determine the critical multiplicity by adjusting $m_c$ until the exponent $\gamma$ in Equation (\ref{eq:lgtM2}) was the same as that in both the gas and liquid phases. Different values of $\gamma$ or $\tau$ will be found for the liquid and gas phases \cite{LGT112}. A similar description and deduction for moment analysis can be found in Ref. \cite{LGT113}. It is commented that this procedure is somewhat subjective since $\gamma$ depends quite strongly on the rang of $\zeta$ selected \cite{LGT112}. Many groups performed the extraction of critical exponents and studied the critical behavior in finite-size system \cite{LGT112,LGT111,LGT114,LGT115,LGT116}, while these results raised controversial debates and makes it an open question for the exponents to indicate the true phase change \cite{LGT118,LGT119,LGT120}.

To find a proper indicator for the liquid-gas phase transition phenomenon, which is not only necessary but also meaningful to guide the experimental analysis and theoretical predictions, Y.G. Ma introduced the information entropy $H$ and the Zipf's law into the diagnosis in 1999 \cite{YG99PRL}. In heavy-ion collisions, by defining $p_i$ as the event probability of having $i$ particles produced, i.e., $\{p_i\}$ is the normalized probability distribution of total multiplicity, and the sum is taken over the whole $\{p_i\}$, the information entropy is constructed as,
\be
H=-\sum_{i}p_i\ln(p_i), \label{eq:EntropyDefHYMa}
\ee
in which $\sum_{i}p_i =$ 1 should be fulfilled. The emphasis is on the event space rather than the phase space, which should be called as the ``event entropy'' or ``multiplicity entropy''. The Zipf's law, which has been known as a statistical phenomenon concerning the relation between the English words and their frequency used in literature in the field of linguistics \cite{LGT121}, was verified to also exist in heavy-ion collisions and be an evidence to characterizing the liquid-gas phase transition.

The isospin dependent lattice gas model (LGM) and the molecular dynamical (MD) model were used to simulate the disassembly of the medium size nucleus $^{129}$Xe. The LGM was proposed by Lee and Yang \cite{LYLGM}, which has been applied to the microscopic nuclear system in the grand canonical ensemble with a sampling of the canonical ensemble \cite{GuptaLGM} and in mean field approximation \cite{LGMMF}. We would not introduce the lattice gas model in this review for the reason it has been incorporated in canonical models. The interested readers can refer to Refs. \cite{YG99PRL,LGTMD01MaJPG} for a brief introduction of the isospin dependent lattice gas model. Three different freeze-out densities $\rho_f$ of the disassembling system were taken in the LGM simulation, i.e., $\rho_f=$ 0.18, 0.38, and 0.60$\rho_0$ ($\rho_0$ is the normal nuclear density), beyond which nucleons are too far apart to interact. In LGM, $\rho_f =$ 0.18$\rho_0$ and 0.60$\rho_0$ correspond to $9^3$ and $6^3$ cubic lattice, respectively. $\rho_f=$ 0.38$\rho_0$ corresponds to $7^3$ cubic lattice since in most cases it is chosen, which corresponds to the experimental data between 0.3$\rho_0$ and 0.4$\rho_0$ \cite{LGT122,LGT123,LGT124}. For the MD simulation, 0.38$\rho_0$ was taken in the situations of with/without the Coulomb interactions.

\begin{figure}[htbp]
\begin{center}
\begin{minipage}[t]{8.5cm}
\epsfig{file=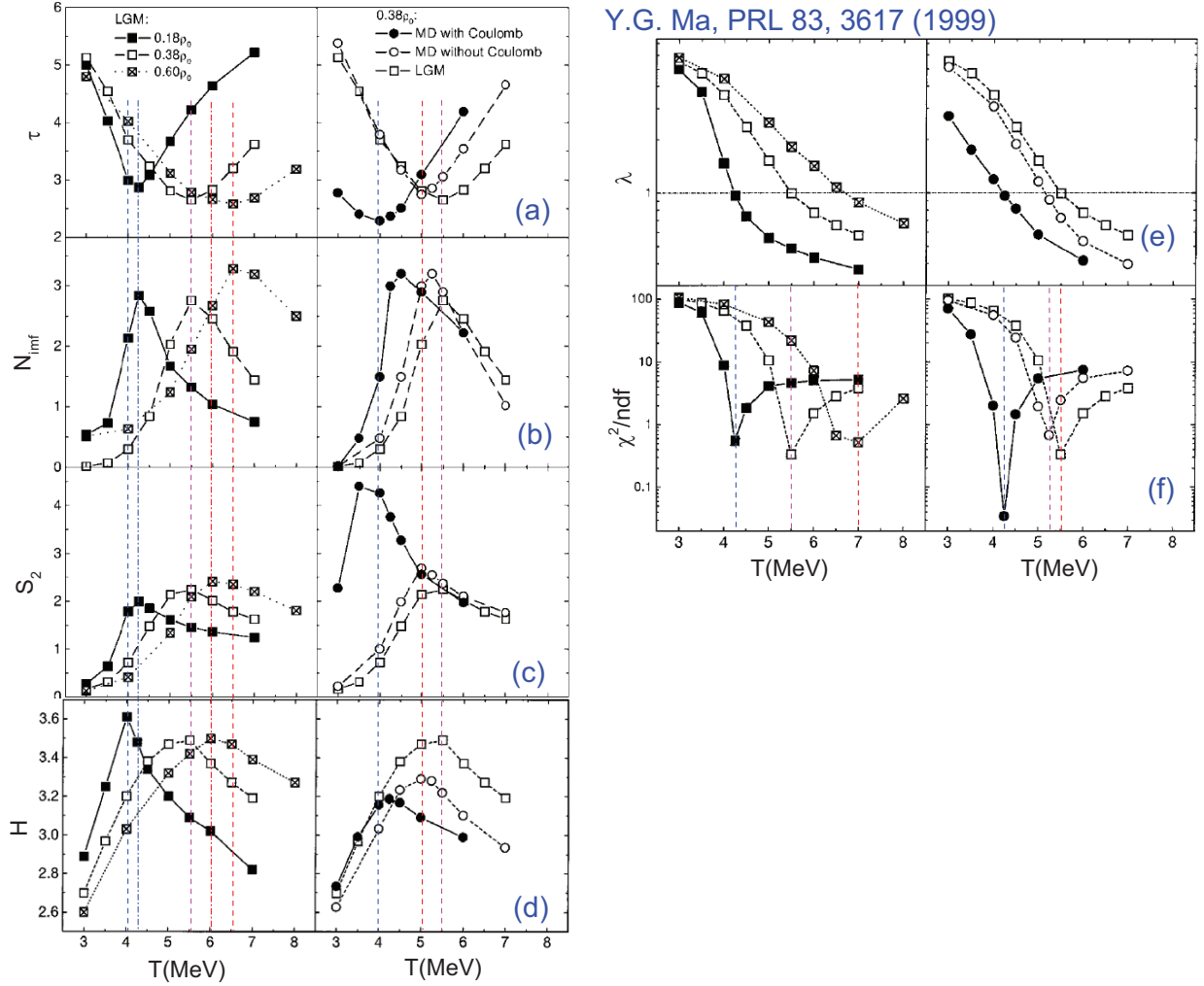,scale=0.75}
\end{minipage}
\begin{minipage}[t]{16.5 cm}
\caption{(Color online) Possible indicators for liquid-gas phase transition in the $^{129}$Xe disassembly simulated by the isospin dependent lattice gas model (LGM) and the molecular dynamical (MD) models. Row (a): critical exponent $\tau$, (b) Multiplicity for intermediate mass fragment $N_{imf}$; (c) Second momentum of cluster distribution $S_2$; (d) Shannon information entropy $H$; (e) Parameter ($\lambda$) fitted in Zipf's law of fragments $A_n\propto n^{-\lambda}$; and (f) $\chi^2/ndf$ with the fit of Zpif's law. The lines in the color are drawn to shown the extreme values of the distributions. The original figures are taken from \cite{YG99PRL}.
\label{LGTygma99prl}}
\end{minipage}
\end{center}
\end{figure}

The critical exponent $\tau$, the second moment of the cluster distribution $S_2$, and the multiplicity of intermediate mass fragments $N_{imf}$ for the disassembly of $^{129}$Xe are plotted as a function of temperature in Figure \ref{LGTygma99prl} (a), (b), and (c), respectively. The information $H$ is plotted as a function of temperature in Figure \ref{LGTygma99prl} (d). For the LGM simulations, the extreme values of $\tau$, $N_{imf}$ and $S_2$ occur at the same temperature for the three freeze-out densities selected. Comparing to the MD simulations at the fixed freeze-out density $\rho_f=0.38\rho_0$, the extreme values occur at different temperatures. Similar distributions of $H$ to chose of $\tau$, $N_{imf}$ and $S_2$ are found. The extreme values of $H$ are clearly shown in the correlation between $H$ and $T$. The extreme values of $H$ are consist to those of $\tau$, $N_{imf}$ and $S_2$, which indicates that the information entropy ought to be a good diagnosis of phase transition. A detailed discussion and more information for this work can be found in Ref. \cite{LGTMD01MaJPG}.

The Zipf's law in the nuclear fragment distribution was firstly investigated by Y.G. Ma \cite{LGTMD01MaJPG}. The size of a fragment and its rank are described by $A_n=c/n$ ($n=$ 1, 2, 3, ...), in which $A_n$ is the mass of rank $n$ in a mass list with the clusters being ordered by decreasing size. Assuming $A_n\propto n^{-\lambda}$ ($\lambda$ is the slope parameter), $\lambda$ can be fitted for fragments at different temperatures (see Figure \ref{LGTygma99prl} (e)). The value of $\lambda$ is found to decrease with temperature, indicating that the difference of mass between the different fragment ranks is becoming smaller. In particular, the Zipf's law is satisfied for $A_n \propto n^{-1}$ when $\lambda\sim-1$ (see Ref. \cite{LGT126}). The temperatures having Zipf's law are also consistent with the transition temperatures extracted from the extreme values of (some) observables shown in Figure \ref{LGTygma99prl} (a)-(d). Further analysis on the Zipf's law for fragments are performed by extracting the truth of hypothesis of $\chi^2$ test. The $\chi^2/ndf$ for the $A_n-n$ relations at different $T$ for different cases are plotted in Figure \ref{LGTygma99prl} (f). The minima of $\chi^2/ndf$ around the respective transition temperature further support the Zipf's law of fragment distribution, which indicates that a liquid-gas phase transition happens. The verification of Zipf's law in experimental analysis can be found in Refs. \cite{LGT116,LGT117}, in which the critical behavior in light Au-like nuclei has been proven in data from TAMU (Texas A$\&$M Univ.) NIMROD (Neutron Ion Multidetector for Reaction Oriented Dynamics) and beams from TAMU K500 super-conducting cyclotron. The Zipf's law was further proven in the multifragment emission in the Pb-Pb and Pb-Plastic collsions of the EMU13 CERN experiments \cite{zipfprven1,zipfprven2}.

For the system, the maximum of $H$ reflects the largest fluctuation of the multiplicity probability distribution in the phase transition point. It is thus difficult to predict how many cluster can be produced in each event space, i.e., the disorder of information is the largest. When using the information entropy to indicate the disorder of a system, the larger the dispersal of multiplicity probability distribution, the higher the information, and then the higher disorder of the system in the event topology. From the statistical point of view, the Zipf's law is related to the critical behavior or self-organized criticality \cite{LGT108,LGT127}. At the liquid-gas phase transition point, the information entropy of multiplicity distribution reaches the maximum, and indicates that the system has the largest fluctuation/stochasticity/chaoricity in the event space at this time.

Balenzuela and Dorso further studied the derivative of the information entropy, together with the other indicators for liquid-gas phase transition in fragmentation system \cite{LGT128}. The generalized R\'{e}nyi entropies were introduced \cite{LGT129,RenyiEntrpy},
\bea\label{Infhighorder}
H_q(S)=\left\{
      \begin{array}{ll}
        -\sum_{i}p_{i}^{n}\ln p_{i}, & q=1  \\
        -\frac{1}{q-1}\ln(\sum_{i}^{n}p_{i}^q), & q > 0, q\neq 1
      \end{array}
    \right.
\eea
When $q =$ 1, $H = H(S)$ which is the Shannon information entropy. In their conclusions, they claimed that the information entropy $S_1$ cannot be the signal for the second order phase transition (this maybe due to that theoretically the finite system and Coulomb effect in heavy-ion collisions can only show the first-order phase transition \cite{FntMall15}). The difference between the $q-$th order of Renyi entropy and the traditional information entropy (i.e. the first order of Renyi entropy) is found just to be a $q$-dependent constant but which is very sensitive to the form of probability distribution \cite{SclBrk02}. Lukierska-Walasek and Toptlski studied the link between the Zipf's law and statistical distributions for the Fortuin-Kasteleyn clusters in the Ising as well as the Potts models \cite{Zipflinksstat}. Conclusions are also drawn that the Zipf's law can be a criterion of a phase transition, but it does not determine the order of the phase transition.

\begin{figure}[htbp]
\begin{center}
\begin{minipage}[t]{15.5cm}
\epsfig{file=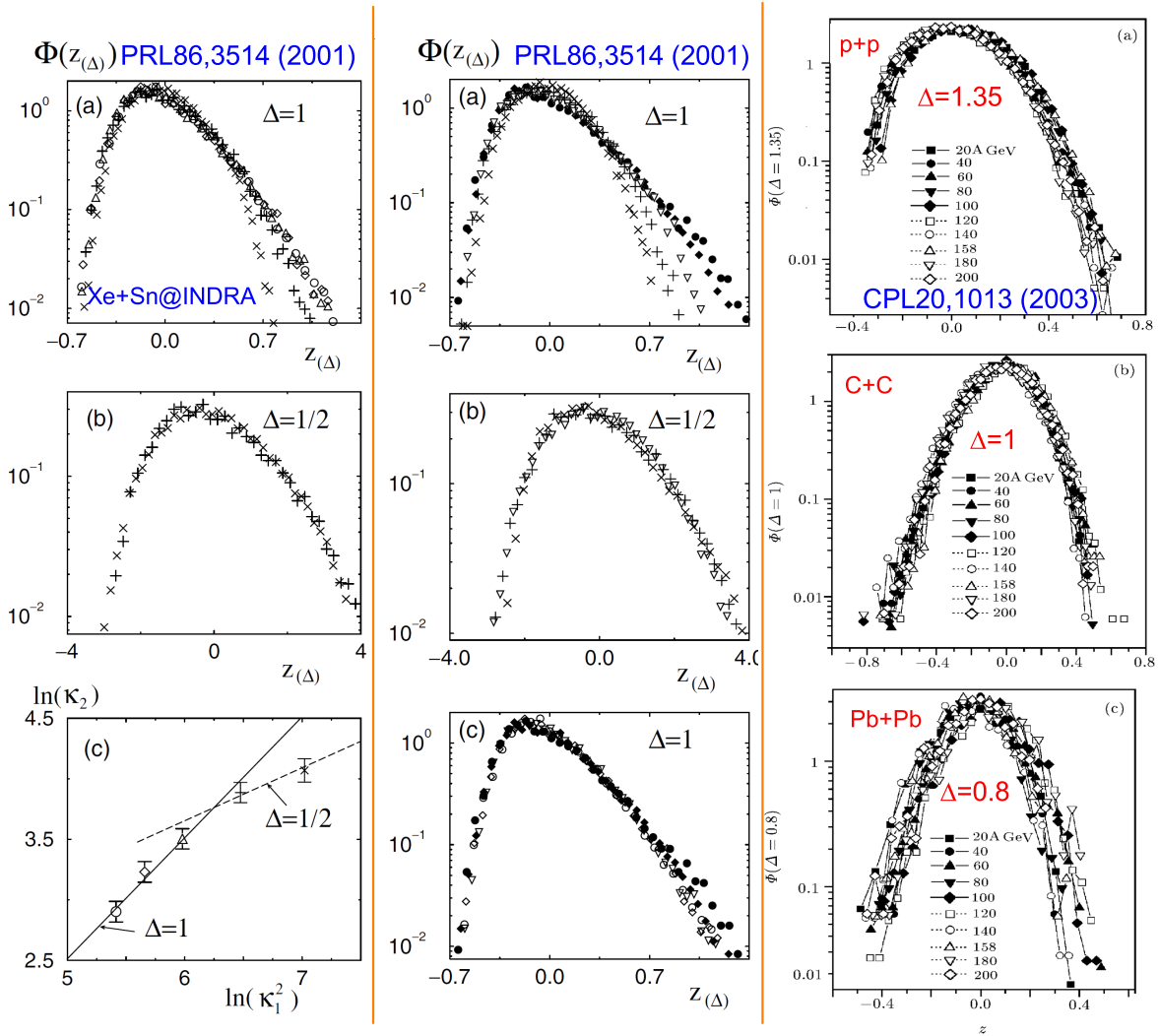,scale=1}
\end{minipage}
\begin{minipage}[t]{16.5 cm}
\caption{(Color online) The $\Delta-$Scaling phenomena in heavy-ion collisions. The left and middle columns: Different characteristics of the largest fragment change probability distributions $P_{N}[m]$ for central Xe + Sn collisions for the INDRA data \cite{LGT130}. The right column: the $\Delta-$Scaling of (a) $p + p$ collisions, (b) C + C collisions, and (c) Pb + Pb collisions with the incident energy ranging from 20 to 200$A$ GeV \cite{LGT132}.
\label{LGTDeltaSCal}}
\end{minipage}
\end{center}
\end{figure}

\subsubsection{$\Delta-$Scaling Phenomenon as Liquid-Gas Phase Transition Indicator}

The $\Delta-$scaling has also been proposed to indicate the liquid-gas phase transition. The $\Delta-$scaling law happens when two or more probability distributions $P_{N}[m]$ of the stochastic observable $m$ collapse on to a single scaling curve $\phi(z)$ if a new scaling observable is defined by,
\be
z=(m-m^{*})/<m>^{\Delta}.
\ee
The curve is,
\be
<m>^{\Delta}P_{N}[m]=\phi(z)=\phi[(m-m^{*})/<m>^{\Delta}],
\ee
where $\Delta$ is a scaling parameter, $m^{*}$ is the most probable value, and $<m>$ is the average of $m$. When $\Delta=$ 1, it is the first scaling law caused by self-similarity of system,which means that the distributions with different $<m>$ (represented by a new kind of variable $z$) entirely collapse on the same curve. In intermediate energy heavy-ion collisions, Botet and Ploszajczak applied the $\Delta-$scaling law to the INDRA data \cite{LGT130}, which are within the intermediate energy range (25-100 MeV/u Xe + Sn), by using $Z_{max}$ (the maximum of charge in reactions) as an order parameter. They found that the distribution of $Z_{max}$ obeys the $\Delta=$ 1/2 scaling law below 45 MeV/u, while it obeys the $\Delta=$ 1 scaling law above 45 MeV/u (see Figure \ref{LGTDeltaSCal}, the left and middle columns), which indicates that a transition occurs around 45 MeV/u from an order phase to the maximum fluctuation phase (disorder phase) \cite{LGT131,LGT132}.

G.L. Ma \textit{et al} \cite{LGT132} applied the $\Delta-$scaling method to study the $p+p$, C + C, and Pb + Pb collisions within the incident energy range of 20-200 GeV/u with the help of the LUCIAE 3.0 \cite{LGT133}, and found that the $\Delta-$scaling phenomena in the ultra-relativistic nucleus-nucleus collisions. Moreover, they introduced the Shannon information entropy for continuum in the analysis by defining,
\be
H_{direct}=-\int P(m)\ln P(m)dm, \label{eq:hdirect}
\ee
where $\int P(m)dm=1$ should be fulfilled. The quality of fluctuation of the system will depend on the distribution $P_N[m]$. Assuming that $P_N[m]$ is a Gaussion distribution, one has,
\be
H_{Gauss} = \ln\sigma + [1 + \ln(2\pi)]/2 \approx \ln\sigma+1.419.  \label{eq:hgauss}
\ee

\begin{figure}[htbp]
\begin{center}
\begin{minipage}[t]{11.5cm}
\epsfig{file=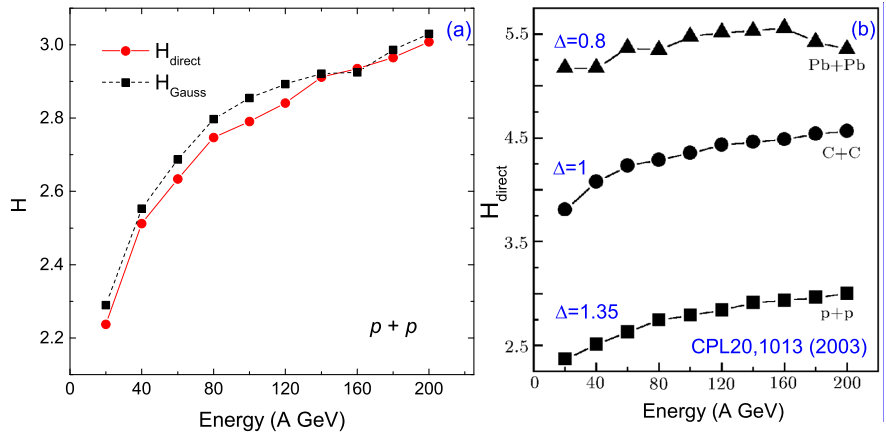,scale=0.6}
\end{minipage}
\begin{minipage}[t]{16.5 cm}
\caption{(Color online) (a) Comparison between $H_{direct}$ and $H_{Gauss}$ for $p + p$ reaction simulated using the LUCIAE 3.0 (data taken from Table 1 in \cite{LGT132}). (b) $H_{direct}$ for the $p + p$, C + C, and Pb + Pb collisions with $\Delta=$ 1.35, 1, and 0.8, respectively, in which the incident energy ranging from 20 to 200$A$ GeV \cite{LGT132}.
\label{LGTDeltaSCHdir}}
\end{minipage}
\end{center}
\end{figure}

The results for $H_{direct}$ and $H_{Gauss}$ have been compared, see Figure \ref{LGTDeltaSCHdir} (a), for which $H_{Gauss}$ is found to slightly larger than $H_{direct}$. The information entropy $H_{direct}$ calculated from the charged multiplicity distributions increases with the beam energy and with the colliding system size monotonically. $H_{direct}$ shows no discontinuity with beam energy [see Figure \ref{LGTDeltaSCHdir} (b)], which is expected since LUCIAE has no change of particle production dynamics, while they are associated with a phase transition in the simulated data. It is expected that the $\Delta-$scaling and entropy variable can be valuable tools to search for possible discontinuities in nucleus-nucleus collisions with the onset of a QCD phase transition. More work on the $\Delta-$scaling in the simulated $p + p$, C + C, and Pb + Pb collisions with LUCIAE 3.0 can be found in Ref. \cite{LGT134}. The recent results from RHIC Beam Energy Scan program \cite{Arxiv10072613} show a non-monotonic behaviour of the energy dependence of net-proton kurtosis \cite{XuN2017NST}, which may be related to the existence of QCD critical point \cite{QCD99CrtP}. It will be interesting to preform the $\Delta$-scaling measurement in the RHIC-BES program, because the QCD phase transition or critical fluctuations could violate the $\Delta$-scaling of probability distribution of some observables such as net-baryon number and harmonic flows \cite{SclBrk02}.

\subsubsection{Isobaric Scaling Phenomenon}

The liquid-gas phase transition is denoted by the fast and violent emission of light particles, i.e., the charged droplet is turned into gas state in which the small clusters are freed from the system by collision. As we have described in previous sections, the process of heavy-ion collisions can be divided into the collision stage and the inflation stage denoted by de-excitation of hot fragment. In the collision stage, nuclear matter of supra-saturation density can be produced due to the compression between the projectile and target nuclei. While in the inflation stage, sub-saturation nuclear matter is formed. It is heavy-ion collisions that provide the important laboratory to study nuclear matters beyond the normal saturation density. Besides the light particles, isotopes heavier than $\alpha$ are measurable. The fragments with $Z >$ 2 which are called as the intermediate mass fragments (IMFs). The IMFs play important roles in investigating the nuclear properties. The nuclear matter and nuclear symmetry energy are of the most important problems in heavy-ion collision physics \cite{IBD135}. From the study of the isospin physics via heavy-ion collisions, with neutron-rich, stable and/or radioactive nuclei, the isospin dependence of in-medium nuclear effective interactions and equation of state of neutron-rich nuclear matter can be studied. Observables constructed from $\pi$, $K$, proton to heavier isotopes are used to study the nuclear symmetry energy and nuclear densities (see reviews \cite{IBD135,IBD136}).

Exotic structures formed in nuclei, such as the halo or skin structures, exist in rare nuclei who have large asymmetry. The very proton-rich or neutron-rich isotopes are important to test the extreme nuclear physics both experimentally and theoretically. The neutron-rich nucleus having large asymmetry are important to both nuclear physics and astrophysics. The new generation of radioactive ion beam (RIB) facilities provide the opportunities to study the asymmetric nuclear matter and nuclear near the drip lines. The very neutron-rich nuclei are usually on the main task of the new generation of RIB facilities \cite{IBD137}. For example, it is one of the main scientific goals for FRIB to investigate the neutron-skin thickness of $^{208}$Pb and $^{40}$Ca, which has been listed in the \textit{2015 long range plan for nuclear science} of USA \cite{IBD138}.

We will not describe the methods to extract nuclear symmetry energy, or neutron-skin thickness since they have been reviewed in \cite{IBD135,IBD136}. Only the methods related this review will be introduced. In heavy-ion collisions, the fragment ratio, from isotopic ratio in single or comparable reactions are used to extract the nuclear symmetry energy for neutron-rich matter at subsaturation densities, which are formed in the freeze-out stage of reaction. The fragments in reactions estimate the correlation between the isotopic ratio and nuclear density. The first experimental investigation was done by Xu \textit{et al} via the isotopic scaling method \cite{IBD139}, which makes the ratio between the isotopic yield in the neutron-rich system and the reference system. The isotopes ranging from $Z=$ 1 to 10 have been identified in the 50 MeV/u $^{112}$Sn + $^{112}$Sn, $^{112}$Sn + $^{124}$Sn, $^{124}$Sn + $^{112}$Sn, and $^{124}$Sn + $^{124}$Sn reactions. The isotopic ratio between the reactions are found to be,
\be
\frac{M_{obs}(N_i,Z_i)}{M_{obs}^{112}(N_i,Z_i)}
=C(\frac{\rho_p}{\rho_p^{112}})^{Z}
(\frac{\rho_n}{\rho_n^{112}})^{N}
=C(\hat{\rho}_{p})^{Z}(\hat{\rho}_{n})^{N}, \label{eq:isoscalingRatio}
\ee
in which $M_{obs}$ is the multiplicity for an isotope with neutron number $N$ and proton number $Z$ in its $i-$th state. The subindex 112 represent the $^{112}$Sn + $^{112}$Sn reaction system. According to the equilibrium formalism \cite{TLGT94,IBD140},
\be
M_{obs} \propto V\rho_{n}^{N}\rho_{p}^{Z}\widetilde{
N,Z}f_{N,Z}(T)\mbox{exp}[B(N,Z)/T],
\ee
where $V$ is the volume, $\rho_n = M_{i}(1,0)/V \propto \mbox{exp}(\mu_n/T)$, and $\rho_p = M_{i}(1,0)/V \propto \mbox{exp}(\mu_p/T)$ are the primary free neutron and proton densities, respectively.  $\mu_{n}$ and $\mu_{p}$ are chemical potential of neutrons and protons, which depend on the temperature and density of the system \cite{FntMall15}; $\widetilde{N,Z} = \sum_{i,stable}(2j_i+1)\mbox{exp}(-E_{i}^{*}/T)$ is the intrinsic partition function summering over the particle stable states of the fragment; $B(N,Z)$, $J_i$, and $E_{i}^{*}$ are the ground state binding energy, spin, and excitation energy of the isotope in the $i-$th state, and $T$ is the temperature. In Equation (\ref{eq:isoscalingRatio}), the difference between the temperature of fragment in two reactions is assumed to be very small and negligible. Meanwhile, the sequential decay of  primary fragments, which is represented by the multiplicative factor $f_{N,Z}(T)$, is also assumed to influence the result obtained by the final fragment very small. The isotopic ratio, isotonic ratio, and isobaric ratio have been analyzed. The isotopic ratios for $Z=$ 3 to 9 are shown to be described by a simple function $y = 1.42^{N}\times 0.68^{Z}$ for $\frac{^{124}Sn}{^{112}Sn}$ according to Equation (\ref{eq:isoscalingRatio}). An isoscaling law is thus indicated by the isotopic or isotonic yield ratios in the form of \cite{Ma14CTP},
\be
R_{21}(N,Z) = Y_{2}(N,Z)/Y_{1}(N,Z)=C\mbox{exp}(\alpha N + \beta Z), \label{eq:isoscaling}
\ee
with $\alpha$ and $\beta$ are isoscaling parameters, which will be explained later. One can rewrite Equation (\ref{eq:isoscaling}) as,
\be
\ln R_{21}(N,Z) = \ln C + N\alpha + Z \beta. \label{eq:isoscalingln}
\ee

In the analysis, if one uses the isotopic ratio to investigate the isoscaling phenomenon, Equation (\ref{eq:isoscalingln}) becomes to,
\be
\ln R_{21}(N,Z) = \ln C + N\alpha,   \label{eq:isoscalinglnZ}
\ee
and for isotonic ratio,
\be
\ln R_{21}(N,Z) = \ln C + Z\beta. \label{eq:isoscalinglnN}
\ee

\begin{figure}[htbp]
\begin{center}
\begin{minipage}[t]{14cm}
\epsfig{file=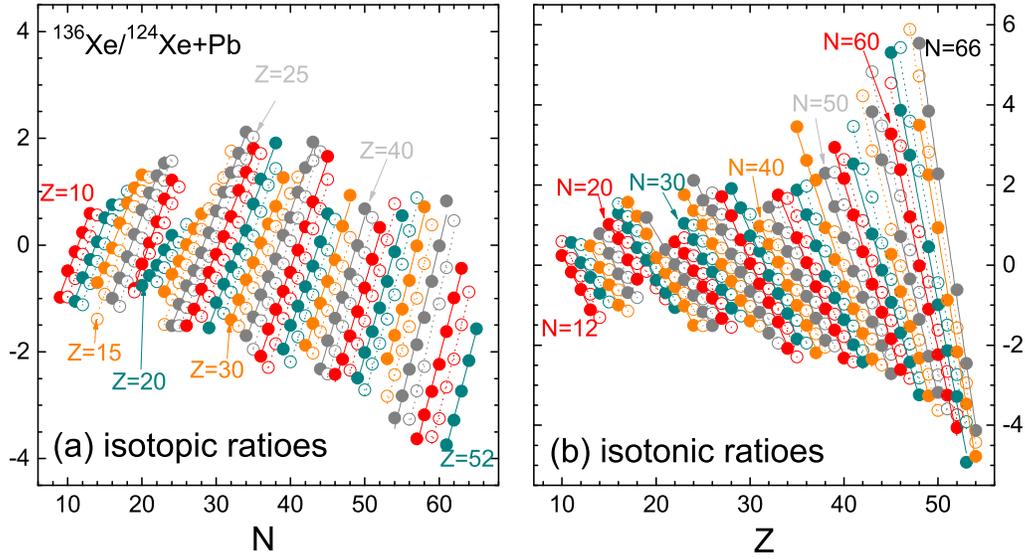,scale=0.7}
\end{minipage}
\begin{minipage}[t]{16.5 cm}
\caption{(Color online) Typical isoscaling phenomena in the measured fragments of the 1 GeV/u $^{136}$Xe/$^{124}$Xe+ Pb projectile fragmentation reactions. (a) The isotopic scaling for isotopes from $Z=$ 10 to 52. (b) The isotonic scaling for isotones from $Z=$ 12 to 66. The measured cross sections of fragments are taken from Ref. \cite{IBD143}.
\label{IBDisoscaling}}
\end{minipage}
\end{center}
\end{figure}

By fitting the correlation between $\ln R_{21}(N,Z)$ and $N$ or $Z$, $\alpha$ or $\beta$ can be determined. A typical isoscaling plot is shown in Figure \ref{IBDisoscaling}, in which a good isotopic scaling [see panel (a)] and isotonic scaling [see panel (b)] can be both found \cite{IBD142} for the measured fragments in the 1 GeV/u $^{136}$Xe/$^{124}$Xe + Pb projectile fragmentation reactions (the measured cross sections for fragments are reported in Ref. \cite{IBD143}).

To describe the isoscaling phenomena, one can rewrite the yield of fragment according to Refs. \cite{TLGT94,IBD140} as,
\be
Y_{i}(N,Z) = F_{i}(N,Z)\mbox{exp}[B(N,Z)/T]
\mbox{exp}[(N\mu_{in}+Z\mu_{ip})/T], \label{eq:fragyield}
\ee
where $F_{i}(N,Z)$ includes the information about the secondary decay of the possible primary fragment. The subindex $i$ denotes the reaction system. One can then define the isotopic ratio for fragments in two reactions \cite{IBD141},
\be
\frac{Y_{2}(N,Z)}{Y_{1}(N,Z)}
=\frac{F_{2}(N,Z)}{F_{1}(N,Z)}
\mbox{exp}[N(\mu_2-\mu_1)/T+ZN(\mu_2-\mu_1)/T],
\ee
$\frac{F_{2}(N,Z)}{F_{1}(N,Z)}=1$ can be assumed since $F(N,Z)$ in the two reactions are assumed as the same. $T$ of the fragment in the two reactions are also assumed to be the same, which results in the cancellation of the binding energy term. Two isoscaling parameters can be related to the density of protons and neutrons as,
\bea
&\hat{\rho}_{n}=\mbox{exp}(\Delta\mu_{n}/T)
=\mbox{exp}(\alpha), \\
&\hat{\rho}_{p}=\mbox{exp}(\Delta\mu_{p}/T)
=\mbox{exp}(\beta).
\eea
Or,
\bea
&\alpha\equiv(\Delta\mu_{n}/T)=(\mu_{n2}-\mu_{n1})/T=\ln\hat{\rho}_{n} , \\
&\beta\equiv(\Delta\mu_{p}/T)=(\mu_{p2}-\mu_{p1})/T=\ln\hat{\rho}_{p} ,
\eea
in which $\Delta\mu_{n}$ and $\Delta\mu_{p}$  are the difference in the neutron and proton chemical potentials between the two reactions; $\hat{\rho}_{n}=\rho_{n2}/\rho_{n1}$ and $\hat{\rho}_{p}=\rho_{p2}/\rho_{p1}$ are the relative ratios of the free neutrons and protons densities in the two reactions.

The isoscaling phenomenon has been found in a variety of reactions over a wide range of energy. Once one know $\mu_n$ and $\mu_p$, the nuclear density will be extracted \cite{IBD156}. This is very important to study the neutron-rich matters of its equation of state. The isoscaling parameter $\alpha$ is related to the nuclear symmetry energy by
\be
\alpha=4C_{sym}[(Z_1/A_1)^2-(Z_2/A_2)]/T.  \label{eq:isoalpha1}
\ee
By defining $\delta_i=(N_i-Z_i)/A_i$, one has,
\be
\alpha=2C_{sym}\Delta\delta(1-\bar{\delta}),\label{eq:isoalpha2}
\ee
in which $\Delta\delta=\delta_2-\delta_1$ and $\bar{\delta}=\delta_2+\delta_1$. Equations (\ref{eq:isoalpha1}) and (\ref{eq:isoalpha2}) show how $\alpha$ depends on the asymmetries of the reactions $\delta_i$ \cite{IBD141}. One can also refer to a work on the systematic study of isoscaling both in theories and experiments \cite{IBD141-1}.

Similar to the isoscaling method, the isobaric yield ratio method also connect the liquid-drop model parameters to the ratio of fragment yields from break-up of hot nuclei \cite{ENST03}. The isobaric ratio of the intermediate-mass-fragments is supposed to be used to extract the difference between the neutrons and protons of the systems, by which the density is also linked to the yields or cross sections of fragments. The isobaric ratio provides more cancellation of parameters which influence the cross section of fragment in theory \cite{IBD144}. If one considers a fragment with mass and neutron-excess $(A, I)$ ($I=N-Z$) with free energy $F(A,I)$, the yield in Equation (\ref{eq:fragyield}) can be written in the form of,
\be
Y(A, I)=CA^{\tau}\mbox{exp}\{[F(A,I)+N\mu_{n}+Z\mu_{p}]/T\}. \label{eq:yieldisobar}
\ee
This corresponds to the canonical ensembles theory within the grand-canonical limit \cite{IBD145,IBD146}. The yield ratio for isobars in a reaction is defined as,
\be
\ln R(I+2m,I,A)=\ln [Y(I+2m,A)/Y(I,A)]=(\Delta F+m\Delta\mu_{np})/T,~~~~m=1,2,3,...
\ee
in which $\Delta F = F(I+2m,A)-F(I,A)$, and $\Delta\mu_{np} = \mu_n - \mu_p$. From isobars of mirror nuclei, $\Delta\mu_{np}$ can be extracted directly from the isobaric yield ratio in one reaction \cite{IBD144,IBD147,IBD148,IBD149,IBD1491},
\be
\ln R(1,-1,A) = [\Delta\mu_{np}+2a_c(Z-1)/A^{1/3}]/T,
\ee
in which $a_c$ is the Coulomb energy coefficient. The correlation between the neutron-skin thickness and $\Delta\mu_{np}$ has been discussed in Ref. \cite{IBD149}.

Based on the deductions in Refs. \cite{Ma15PLB,IBD154,IBD142,IBD156,IBD150}, the isobaric ratio difference between two reactions will be,
\bea
\Delta\ln R^{IB}(I+2m,I,A)=&m[(\mu_{n2}-\mu_{n1})/T-\mu_{p2}-\mu_{p1})/T]\\
                =&m(\Delta\mu_{n}/T-\Delta\mu_{p}/T)=m(\alpha-\beta),
\eea
or,
\bea
\Delta\ln R^{IB}(I+2m,I,A)=&m[(\mu_{n2}-\mu_{n1})/T-\mu_{p2}-\mu_{p1})/T]\\
                =&m(\Delta\mu_{np2}/T-\Delta\mu_{np1}/T),
\eea

The influence of neutron-skin thickness of projectile nucleus has been investigated via $\Delta\ln R^{IB}(I+2,I,A)$ in Refs. \cite{IBD152,IBD151}. At the same time, it was first illustrated that $\Delta\ln R^{IB}(I+2,I,A)$ for fragments with different $I$ has a scaling phenomenon, which have very similar distributions and indicate that the fragments carry similar information entropy for the system \cite{Ma15PLB}. A further analysis found that, besides the scaling of $\Delta\ln R^{IB}(I+2,I,A)$, the scaling phenomenon also exists in $\Delta\ln R^{IB}(I+2m,I,A)$, for which the isobars have more larger difference in neutron-excess \cite{IBD154}. The information entropy carried by the fragment is different to the definition for the entropy of the system, i.e., $H = -\sum p_{i}\ln p_i$, which is defined for the event space.

In a reaction system which has multi events $S=\{E_{1}, E_{2},... E_{n}\}$ and the corresponding probabilities
$P=\{p_{i}, p_{2}, ..., p_{n}\}$, one can consider the fragment as $E_{i}$ and the probability $p_{i}$ as its cross section ($\sigma$) \cite{Ma15PLB,IBD154,IBD155}. The information entropy carried by the fragment, which is named as the {\it information uncertainty}, is defined as,
\be
In(E_i)=-\ln p_{i}.
\ee
This definition for a specific fragment is an subcomponent of system information entropy $H$, which is supposed to reflect the properties of the system.

\begin{figure}[htbp]
\begin{center}
\begin{minipage}[t]{14cm}
\epsfig{file=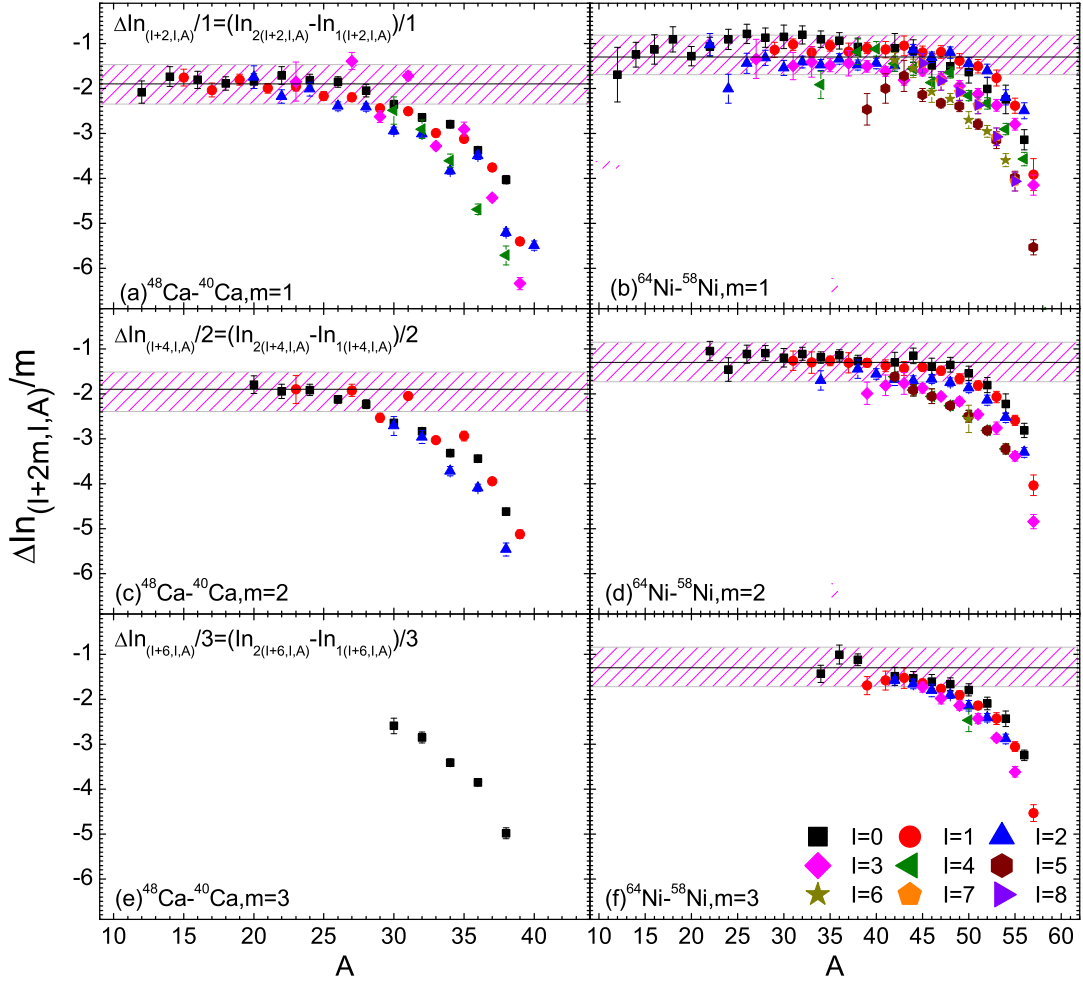,scale=1.1}
\end{minipage}
\begin{minipage}[t]{16.5 cm}
\caption{(Color online) The isobaric information uncertainty phenomenon for the measured fragments in the projectile fragmentation reactions of: (a) the 140 MeV/u $^{48}$Ca/$^{40}$Ca + $^{9}$Be; (b) and $^{64}$Ni/$^{58}$Ni + $^{9}$Be with $m =$ 1 in Equation (\ref{eq:infoScalsimp}). (c) and (d) are the same as (a) and (b) but for $m =$ 2. (e) and (f) are the same as (a) and (b) but for $m =$ 3.
\label{InfoscalingDifmCaNi}}
\end{minipage}
\end{center}
\end{figure}

Before one uses $\sigma(I,A)$ to calculate $In(E_i)$ for a fragment, it should be noticed that for all the fragments the following equation should be physically fulfilled,
\be
\sum_{i=1}^{m}p_{i}=1,
\ee
where $m$ denotes the numbers of fragment species, including the measured and unmeasured ones. One can define the total production cross section of fragments as,
\be
\sigma_{Ft}\equiv\sum_{i=1}^{m}\sigma_{i}(I,A).
\ee
One then has,
\be
In(I,A)=-\ln[\sigma(I,A)/\sigma_{Ft}]=-\ln \sigma(I,A)+const.  \label{eq:infoIndef}
\ee
Then one can define the difference between the information uncertainties of isobars which have neutron-excess difference of $2m$,
\be
\Delta In_{(I+2m, I, A)}=In_{2(I+2m,I,A)}-In_{1(I+2m,I,A)},
~~~~m=1,2,3,...
\ee
Inserting Equation (\ref{eq:yieldisobar}) into (\ref{eq:infoIndef}), which is similar to the definition of the isobaric ratio difference between two reactions, one has,
\be
\Delta In_{(I+2m, I, A)}=m[(\mu_{n1}-\mu_{n2})-(\mu_{p1}-\mu_{p2})]/T,
~~~~m=1,2,3,...\label{eq:infoScalDef}
\ee
which results in a general formula,
\be
\Delta In_{(I + 2m, I, A)}/m=[(\mu_{n1}-\mu_{n2})-(\mu_{p1}-\mu_{p2})]/T,
~~~~m=1,2,3,... \label{eq:infoScalsimp}
\ee
Comparing to $\Delta \ln R^{IB}_{(I + 2m, I, A)}$, it can be easily found that,
\be
\Delta In_{(I + 2m, I, A)}/m=-\Delta \ln R^{IB}_{(I + 2m, I, A)},
~~~~m=1,2,3,... \label{eq:infoScalandIBD}
\ee

\begin{figure}[htbp]
\begin{center}
\begin{minipage}[t]{15cm}
\epsfig{file=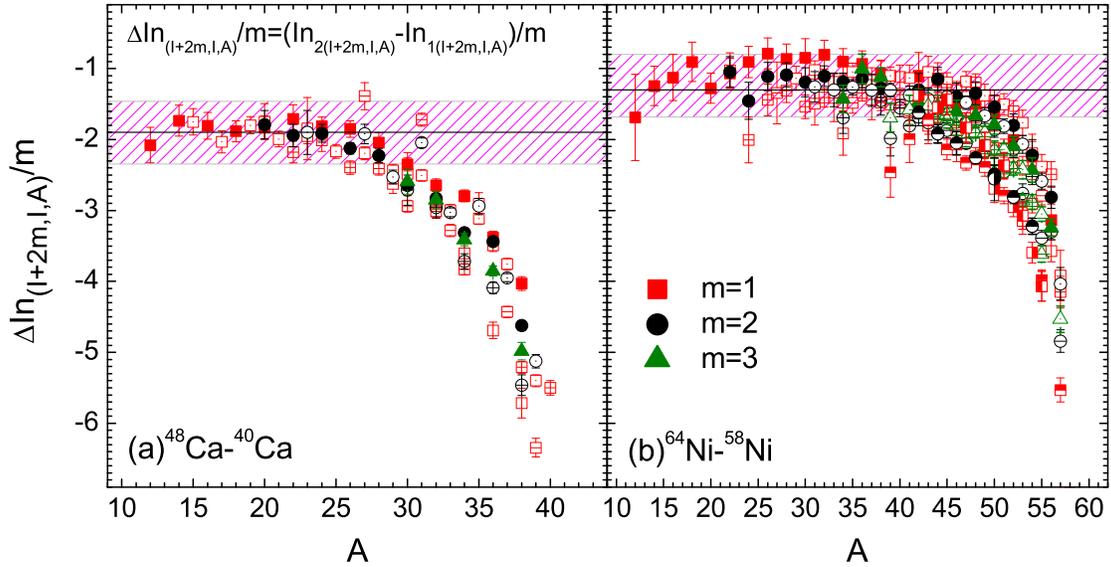,scale=1}
\end{minipage}
\begin{minipage}[t]{16.5 cm}
\caption{(Color online) The scaling phenomena of $\Delta In_{(I + 2m, I, A)}/m$ for fragments with different $m$ in the 140 MeV/u $^{48}$Ca/$^{40}$Ca + $^{9}$Be (a) and $^{64}$Ni/$^{58}$Ni + $^{9}$Be (b) reactions.
\label{InfoscalingCaNi}}
\end{minipage}
\end{center}
\end{figure}

The fragments in the 140 MeV/u $^{40, 48}$Ca + $^{9}$Be and $^{58, 64}$Ni + $^{9}$Be reactions, which have been measured by Mocko \textit{et al} at the National Superconducting Cyclotron Laboratory (NSCL) in Michigan State University \cite{IBD157}, have been adopted to perform the information uncertainty analysis. Two different scaling phenomena are indicated by Equations (\ref{eq:infoScalDef}) and (\ref{eq:infoScalsimp}). The first one is that for fixed $m$, for example $\Delta In_{(I + 2, I, A)}$, should show a scaling phenomena. The $\Delta In_{(I + 2m, I, A)}/m$ for fragments in the $^{48}$Ca/$^{40}$Ca + $^{9}$Be and $^{64}$Ni/$^{58}$Ni + $^{9}$Be reactions with the fixed $m$ are plotted in Figure \ref{InfoscalingDifmCaNi}. For $m=$ 1, 2, 3, the results of $\Delta In_{(I + 2m, I, A)}/m$ show good scaling phenomena for both the $^{48}$Ca/$^{40}$Ca + $^{9}$Be and $^{64}$Ni/$^{58}$Ni + $^{9}$Be reactions. The distribution of $\Delta In_{(I + 2m, I, A)}/m$ is formed by a plateau and a decreasing part. Typically, the plateaus for $^{48}$Ca/$^{40}$Ca and $^{64}$Ni/$^{58}$Ni are at around -1.85$\pm$0.25 and -0.5$\pm$0.25, respectively \cite{IBD154,Ma15PLB}. While those for the $\Delta In_{(I + 4, I, A)}/2$ and $\Delta In_{(I + 6, I, A)}/3$ are similar to those for the $\Delta In_{(I + 2, I, A)}$ \cite{IBD154}. This happens because that the second scaling phenomenon indicated by Equation (\ref{eq:infoScalDef}). The $\Delta In_{(I + 2m, I, A)}/m$ with different $m$ for the $^{48}$Ca/$^{40}$Ca + $^{9}$Be and $^{64}$Ni/$^{58}$Ni + $^{9}$Be reactions are compared in Figure \ref{InfoscalingCaNi}. $\Delta In_{(I + 2m, I, A)}/m$ shows good scaling phenomena for these reactions, which reflects the consistent of distribution for $[(\mu_{n1} - \mu_{n2}) - (\mu_{p1} - \mu_{p2})]/T$.

\begin{figure}[htbp]
\begin{center}
\begin{minipage}[t]{15cm}
\epsfig{file=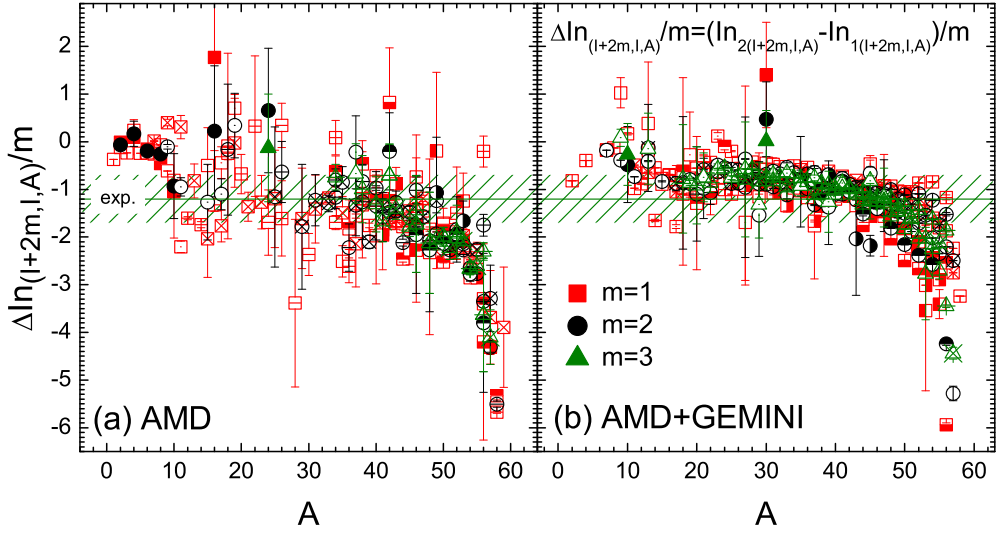,scale=1}
\end{minipage}
\begin{minipage}[t]{16.5 cm}
\caption{(Color online) The scaling phenomena of $\Delta In_{(I + 2m, I, A)}/m$ for fragments with different $m$ in the 140 MeV/u $^{64}$Ni/$^{58}$Ni + $^{9}$Be reactions simulated by the AMD at $t = $ 500 fm/c (a), and the AMD + {\sc gemini} model (b).
\label{InfoscalingCaNiAMD}}
\end{minipage}
\end{center}
\end{figure}

We note that, though Equation (\ref{eq:yieldisobar}) is for the cross section of fragment in the framework of the canonical ensemble theory within the grand-canonical limitation, the scaling phenomena in $\Delta In_{(I + 2m, I, A)}/m$ for isobars with same and different difference in $I$ are discovered in experimental data, which is independent of theoretical models \cite{Ma15PLB,IBD154}. In Figure \ref{InfoscalingCaNiAMD}, the results of $\Delta In_{(I + 2m, I, A)}/m$ for the fragments produced in the 140 MeV/u $^{64}$Ni/$^{58}$Ni + $^{9}$Be reactions simulated by the AMD (+ {\sc gemini}) model are plotted. To simulated the reactions, the extended version of the AMD model (AMD-V) \cite{AMD80} has been used for the reason that it yields good results for intermediate mass fragments \cite{IBD144,Hipse-mocko,IBD154,IBD150,IBD161,IBD162,QiaoTNst}. The soft interaction, i.e., the standard Gogny-g0 interaction \cite{IBD164} was used to govern all the reactions. In the fragment analysis, a coalescence algorithm was adopted with a relatively small coalescence radius $R_c =$ 2.5 fm in the phase space at $t =$ 500 fm/c. The primary fragments reorganized in the phase space of the AMD simulation are allowed to decay by the sequential decay code {\sc gemini} \cite{IBD165}. The fragments after the decaying process are cold fragments, of which the cross sections can be compared to the experimental measured ones. A comparison among the cross sections of fragments by the SAA, {\sc expax2} \cite{IBD167} and {\sc epax3} \cite{IBD168}, as well as the measured results for the 140 MeV/u $^{64}$Ni, $^{58}$Ni + $^{9}$Be reactions can be found in Ref. \cite{IBD163}. Some isobaric yield distributions for the fragments in the simulated reactions can be found in Ref. \cite{IBD166}. In Refs. \cite{IBD163,IBD166}, it can be seen that the cross sections of the primary fragments simulated by the AMD model have a large difference to those of the measured results. It is easy to find in Figure \ref{InfoscalingCaNi} that, a relatively large fluctuation in $\Delta In_{(I + 2m, I, A)}/m$ for the primary fragments obtained by the AMD simulations. While after the decaying process simulation by {\sc gemini}, the scaling of $\Delta In_{(I + 2m, I, A)}/m$ for the cold fragments becomes much better.

The scaling of $\Delta In_{(I + 2m, I, A)}/m$ provides us new ideas to extract the neutron density distribution for neutron-rich nucleus. In Equation (\ref{eq:infoScalandIBD}), for a system with $\mu_{n1}\approx\mu_{p1}$, i.e., a symmetric system with $\rho_{n1} = \rho_{n2}$, one can determine $\mu_{n2}-\mu_{p2}$ from the $\Delta In_{(I + 2m, I, A)}/m$ scaling. Considering that $\rho_{p2}$ can be much easier experimentally extracted by the ways of electron or proton scattering \cite{pscatt17PPNP,escatt17PPNP}, $\mu_{n2}$ and $\rho_{n2}$ can be extracted from the $\Delta In_{(I + 2m, I, A)}/m$ scaling \cite{IBD156}.

\begin{figure}[htbp]
\begin{center}
\begin{minipage}[t]{17cm}
\epsfig{file=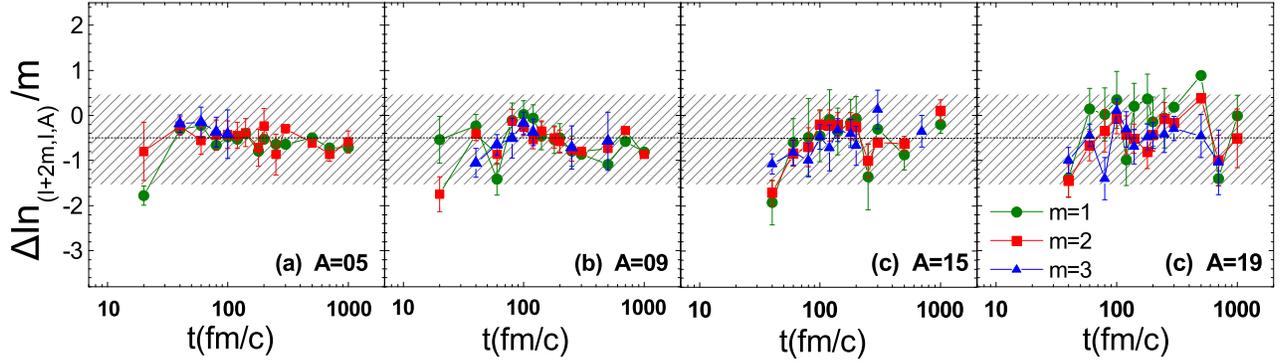,scale=2}
\end{minipage}
\begin{minipage}[t]{16.5 cm}
\caption{(Color online) Preliminary results for time evolution of $\Delta In_{(I + 2m, I, A)}/m$ for isobaric difference with different $m$ in the 140 MeV/u $^{64}$Ni/$^{58}$Ni + $^{9}$Be reactions simulated by the AMD from the colliding times $t =$ 20 to 1000 fm/c. (a), (b), (c), and (d) correspond to $A =$ 5, 9, 15, and 19, respectively.
\label{InIBDEvol}}
\end{minipage}
\end{center}
\end{figure}

The information uncertainty can serve more important roles in the study of the dynamical process of heavy-ion collisions. Though is was shown that $\Delta In_{(I + 2m, I, A)}/m$ for hot fragments in the simulated 140 MeV/u $^{58, 64}$Ni + $^{9}$Be reactions has a scaling phenomenon at $t = $ 500 fm/c, it is at a relatively late time of collisions. One does not know at what time this scaling phenomenon happens. More investigation should be performed. With the help of transport models, one can study the time evolution of the colliding system. A systematic evolution simulations for the 140 MeV/u $^{58, 64}$Ni + $^{9}$Be reactions have been performed using the AMD-V model with the Gogny-g0 interaction, at colliding times $t =$ 20, 40, 60, 80, 100, 120, 140, 180, 200, 250, 300, 500, 700, and 1000 fm/c. Since the phase space at early time collision will be analyzed, a coalescence radius of $R_c =$ 1.5 fm is adopted to reorganize the primary fragments within a range of impact parameters $b =$ 0 to 2 fm. Preliminary results $\Delta In_{(I + 2m, I, A)}/m$ for isobars of $A =$ 5, 9, 15, and 19, with $m =$ 1, 2, and 3, are plotted in Figure \ref{InIBDEvol}. A trend of increasing $\Delta In_{(I + 2m, I, A)}/m$ with $t$ can be seen in the early colliding stage and reaches the maximums, and then decreases to a relatively stable values for $A =$ 5 and 9 isobars. The values of $\Delta In_{(I + 2m, I, A)}/m$ for the $A =$ 5, 9 and 15 isobars are consistent after $t = \sim$ 60 fm/c, showing a good scaling phenomenon. A large fluctuation of $\Delta In_{(I + 2m, I, A)}/m$ is shown for the $A =$ 19 isobars.

\subsection{Prospects}
As a general tool of analyzing information, the Shannon information entropy essentially reveals the quantity of information in a quantity with its distribution, whether in the discrete or continuous shapes. From the point view of configurational entropy, which avoids the \textit{ad hoc} continuum limit of the Shannon information entropy for modes of $n\rightarrow \infty$, can be more properly used in the description of the growth rate of the Shannon information entropy in successively finer discretizations of the space \cite{ComShnn1}. The basic scientific meaning of Shannon information entropy applied in heavy-ion collisions is to indicate the chaoticity of nuclear matter in the colliding nuclear system. The chaoticity analysis based on information entropy, which reflects the order of a colliding system, can serve as indicators for the evolution of nuclear matters. For a wide range of incident energies, i.e., from the intermediate energy to the relativistic energy of AGS and SPS, the scaling of a various physical parameters have been found, which illustrates the order of matters and reflects the consistence of chaoticity in the system.  As shown in this review, besides the liquid-gas phase transition indicated by light particles, the general scaling phenomena ($\Delta$ scaling and isobaric scaling) and in potential their breaking can be indicated from the chaoticity of system. A connection between the ideas of information entropy and the physical quantities of interest provides new insights to the problems concerned. New applications of information entropy and its extended analysis in particle and nuclear physics \cite{Soli01,SclBrk01,SclBrk02} illustrated the entropy scaling in AA collisions at AGS and SPS energies, and suggested new parameters to study the multiparticle production in high energy hadronic and heavy-ion collisions \cite{SclBrk01}. Actually, information entropy has a strong relationship with cumulants which are widely measured in relativistic heavy-ion experiments, e.g. the cumulants of baryon number \cite{Cumulant01} and collective flow \cite{Cumulant02}, since they both characterize the features of probability distribution. Therefore, information entropy is capable of probing the dynamical change during phase transition, which actually results from the change of probability distribution, e.g. from Gaussian to non-Gaussian \cite{SclBrk02}.

A new era of heavy-ion collision physics will be opened by the newly updated/constructed or recently proposed radioactive ion beam (RIB) factories, in which high quality and much more asymmetric nuclear beams can be provided. Nuclear scientists have the opportunity to check the nuclear theories at the extreme conditions, in which more exotic phenomena are hoped.

Compared to the deep development of the information entropy analysis in other scientific areas, the applications of Shannon information entropy analysis in heavy-ion collision physics are much simpler. A deeper study of heavy-ion collisions based on information entropy is urgently needed for the various phenomena, in particular those cannot be directly investigated from physical tools. The configurational entropy is suggested to be a very useful tool to investigate the stability and/or the relative dominance of states for diverse physical system \cite{CEAPP03,KaraCEAPP04EPL}, which has been used in the RHIC and LHC physics. The ideas of configurational entropy can be naturally incorporated in dynamical simulations for the evolution of phase space of systems, and should shed new light in heavy-ion physics. Besides, provided by the advantage of information entropy analysis, the heavy-ion collisions can be analyzed from the point view of information evolution which does not require the system state like in the physical theories. It should be important to develop systematic analysis methods based on information entropy, or to incorporate the information analysis in present theories for heavy-ion collisions.

\section{Summary}
\label{sec:summary}

After the Shannon information entropy theory has been constructed, it has been widely used in a various areas of science with different kind of extensions and developments. In particular, the configurational entropy, which is a good tool to study the dynamical process, has been applied in many disciplines in physics, for example the studies of black hole, scalar glueballs, soliton, $b$-CGC in RHIC and LHC etc. From the content of information entropy, it is illustrated that the nuclear configurational entropy is a way to indicate the onset of the quantum regime and to study the encompasses quantum mechanics fluctuations. As a typical dynamical evolution process, the Shannon information entropy shall benefit the researches in heavy-ion collisions in many ways.

We focus this review on the applications of Shannon information entropy in heavy-ion collisions. The Shannon information entropy has been used both for the discrete and continuous states for branching process and differentiates the QCD and QED models, indicating that the information entropy indices are appropriate parameter to measure the chaotic behavior of particle production, and characterize the degree of fluctuation of parton multiplicity which initiates branching. In the study of liquid-gas phase transition in heavy-ion collisions, the proposed information entropies for light particles in discrete and continuum stochastic observable $m$ both indicate the first-order liquid-gas phase transition. The event information entropy, which is also called as the information uncertainty, has been employed to analyze the information entropy measured from fragments in heavy-ion collisions in neutron-rich systems. The isobaric scaling found in the neutron-rich fragments, and fragments with different difference of neutron-excess reflect the properties of nuclear matter in the systems from the point view of chaoticity, which provides new methods to study the evolution of nuclear matters in heavy-ion collisions.

As a general tool for analyzing the dynamical process of a system, the Shannon information entropy provides methods to study the evolution of colliding nuclear systems in the model-independent manner. For the dynamical process in heavy-ion collision systems, with only local equilibrium state achievable in traditional nuclear reaction theories, the observables for systematic evolution should be carefully constructed. While the typical characteristics of Shannon information entropy analysis methods treat the evolution as an evolution in chaoticity of the system, avoids the requirements of equilibrium state of heavy-ion collisions and bridges the observables in dynamics models and statistical models, and provides new methods to study the time evolution of nuclear matters in the collisions. It is proposed that the information entropy can be included in the simulations of transport models and the evolution of system can be obtained directly. The deep development of information entropy based method and new observables will shed new light to the research of heavy-ion collision physics, which should be very important in the new era of radioactive ion beam characterized by high quality and much more asymmetric nuclear beams.

\section*{Acknowledgement}

We thank Prof. Y. Wei (Southeast Univ., China), S.G. Zhou (ITP, CAS), G.L. Ma (SINAP, CAS), F. Jin (Chongqing Cancer Inst., China) for reading the manuscript. This work is supported by the National Natural Science Foundation of China (grant No. 11421505, 11220101005, U1732135), Major State Basic Research Development Program in China (grant No. 2014CB845401), the Key Research Program of Frontier Sciences of CAS (grant No. QYZDJSSW-SLH002), Natural Science Foundation of Henan Province (grant No. 162300410179), and Henan Normal University for the Excellent Youth (grant No. 154100510007).

\end{document}